\documentclass[12pt]{article}

\usepackage{amsmath,enumerate,amsbsy,amsfonts,amssymb,amscd,graphicx,algorithm,mathabx}
\usepackage[round,authoryear]{natbib}
\usepackage{authblk}
\usepackage{float}
\usepackage{array}
\usepackage{multirow}
\usepackage{lscape} 
\usepackage{xcolor} 

\usepackage{epsfig}

\newtheorem{theorem}{Theorem}

\newtheorem{condition}{Condition}
\newtheorem{lemma}{Lemma}

\newcommand{\blind}{1}
\newcommand{\bOmega}{\boldsymbol \Omega}
\newcommand{\bSigma}{\boldsymbol \Sigma}

\newcommand{\bdelta}{\boldsymbol \delta}

\newcommand{\ba}{{\mathbf a}}

\newcommand{\mR}{{\mathbb R}}

\newcommand{\cU}{{\cal U}}

\newcommand{\cS}{{\cal S}}
\newcommand{\cD}{{\cal D}}
\newcommand{\cH}{{\cal H}}

\newcommand{\cF}{{\cal F}}

\newcommand{\bB}{{\bf B}}

\newcommand{\bK}{{\bf K}}

\newcommand{\bI}{{\bf I}}

\newcommand{\cov}{\text{Cov}}
\newcommand{\vect}{\text{vec}}

\newcommand{\T}{\mbox{$\scriptscriptstyle T$}}

\RequirePackage[colorlinks,citecolor=blue,urlcolor=blue]{hyperref}

\begin{document}
\if1\blind
{
\title{\bf \Large Functional Linear Regression: Dependence and Error Contamination}
\author[1]{Cheng Chen}
\author[2]{Shaojun Guo}
\author[1]{Xinghao Qiao}
\affil[1]{Department of Statistics, London School of Economics, U.K.}
\affil[2]{Institute of Statistics and Big Data, Renmin University of China, P.R. China}
\setcounter{Maxaffil}{0}

\renewcommand\Affilfont{\itshape\small}
\date{\vspace{-5ex}}
\maketitle
} \fi
\if0\blind
{
	\bigskip
	\bigskip
	\bigskip
	\begin{center}
		{\bf \Large  Functional Linear Regression: Dependence and Error Contamination
		}
	\end{center}
	\medskip
} \fi

\bigskip	
	\begin{abstract}
	Functional linear regression is an important topic in functional data analysis. It is commonly assumed that samples of the functional predictor are independent realizations of an underlying stochastic process, and are observed over a grid of points contaminated by i.i.d. measurement errors. In practice, however, the dynamical dependence across different curves may exist and the parametric assumption on the error covariance structure could be unrealistic. In this paper, we consider functional linear regression with serially dependent observations of the functional predictor, when the contamination of the predictor by the white noise is genuinely functional with fully nonparametric covariance structure. Inspired by the fact that the autocovariance function of observed functional predictors automatically filters out the impact from the unobservable noise term, we propose a novel autocovariance-based generalized method-of-moments estimate of the slope function. We also develop a nonparametric smoothing approach to handle the scenario of partially observed functional predictors. The asymptotic properties of the resulting estimators under different scenarios are established. Finally, we demonstrate that our proposed method significantly outperforms possible competing methods through an extensive set of simulations and an analysis of a public financial dataset.
    \end{abstract}
	
	\noindent {\small{\it Some key words}: Autocovariance; Eigenanalysis; Errors-in-predictors; Functional linear regression; Generalized method-of-moments; Local linear smoothing.}
	
\section{Introduction}
\label{sec.intro}
In functional data analysis, the linear regression problem depicting the linear relationship between a functional predictor and either a scalar or functional response, has recently received a great deal of attention. See \cite{Bramsay2005} for a thorough discussion of the issues involved with fitting such data. For examples of recent research on functional linear models, see \cite{yao2005,hall2007,crambes2009,cho2013,chak2017} and the references therein. We refer to \cite{morris2015} for an extensive review on recent developments for functional regression. 

In functional regression literature, one typical assumption is to model observed functional predictors, denoted by $X_1(\cdot), \dots, X_n(\cdot),$ as independent realizations of an underlying stochastic process. However, curves can also arise from segments of consecutive measurements over time. Examples include daily curves of financial transaction data \cite[]{horvath2014}, intraday electricity load curves \cite[]{cho2013} and daily pollution curves \cite[]{aue2015}. Such type of curves, also named as curve time series, violates the independence assumption, in the sense that the dynamical dependence across different curves exists. The other key assumption treats the functional predictor as being either fully observed \cite[]{hall2007} or incompletely observed, with measurement error, at a grid of time points \cite[]{crambes2009}.  In the latter case, errors associated with distinct observation points are assumed to be i.i.d., where the corresponding covariance function for the error process is diagonal with constant diagonal components. In the curve time series setting, $X_t(\cdot)$ are often recorded at discrete points and are subject to dependent and heteroskedastic errors. Hence, the resulting error covariance matrix would be more nonparametric with varying diagonal entries and nonzero off-diagonal entries.


In this paper, we consider the functional linear regression in a time series context, which involves serially dependent observations of the functional predictor contaminated by genuinely functional errors corresponding to a fully nonparametric covariance structure. We assume that the observed erroneous predictors, which we denote by $W_1(\cdot), \dots, W_n(\cdot),$ are defined on a compact interval $\cU$ and are subject to errors in the form of 
\begin{equation}
\label{curve.decomp}
W_t(u)=X_t(u)+e_t(u), ~~ u \in \cU,
\end{equation}
where the error process $\{e_t(\cdot), t = 1, 2,\ldots\}$  is a sequence of white noise such that $E\{e_t(u)\}=0$ for all $t$ and $\cov\{e_t(u),e_{s}(v)\}=0$ for any $(u,v) \in {\cU}^2$ provided $t \neq s.$
We also assume that $X_t(\cdot)$ and $e_t(\cdot)$ are uncorrelated and correspond to unobservable signal and noise components, respectively. The error contamination model in (\ref{curve.decomp}) was also considered in \cite{bathia2010}. To fit the functional regression model, the conventional {\it least square} (LS) approach \cite[]{hall2007} relies on the sample covariance function of $W_t(\cdot)$, which is not a consistent estimator for the true covariance function of $X_t(\cdot)$, thus failing to account for the contamination that can result in substantial estimation bias. One can possibly implement the LS method in the resulting multiple linear regression after performing dimension reduction for $W_t(\cdot)$ to identify the dimensionality of $X_t(\cdot)$ \cite[]{bathia2010}. However, this approach still suffers from unavoidable uncertainty due to $e_t(\cdot),$ while the inconsistency has been demonstrated by our simulations. Inspired from a simple fact that $\cov\{W_t(u),W_{t+k}(v)\}=\cov\{X_t(u),X_{t+k}(v)\}$ for any $k \neq 0,$ which indicates that the impact from the unobservable noise term can be automatically eliminated, we develop an {\it autocovariance-based generalized method-of-moments} (AGMM) estimator for the slope function. This procedure makes the good use of the serial dependence information, which is the most relevant in the context of time series modelling.

To tackle the problem we consider, the conventional LS approach is not directly applicable in the sense that one cannot separate $X_t(\cdot)$ from $W_t(\cdot)$ in equation (\ref{curve.decomp}). This difficulty was resolved in \cite{hall2006} under the restrictive ``low noise" setting, which assumes that the noise $e_t(\cdot)$ goes to zero as $n$ grows to infinity. The recent work by \cite{chak2017} implements the regression calibration approach combined with the low rank matrix completion technique to separate $X_t(\cdot)$ from $W_t(\cdot).$ Their approach relies on the identifiability result that, provided real analytic and banded covariance functions for $X_t(\cdot)$ and $e_t(\cdot),$ respectively, the corresponding two covariance functions are identifiable \cite[]{descary2019}. However, all the aforementioned methods are developed under the critical independence assumption, which would be inappropriate for the setting that $W_1(\cdot), \dots, W_n(\cdot)$ are serially dependent. 


The proposed AGMM method has four main advantages. 
First, it can handle regression with serially dependent observations of the functional predictor. The existence of dynamical dependence across different curves makes our problem tractable and facilitates the development of AGMM. 
Second, without placing any parametric assumption on the covariance structure of the error process, it relies on the autocovariance function to get rid of the effect from the genuinely functional error. Interestingly, it turns out that the operator in AGMM defined based on the autocovariance function of the curve process is identical to the nonnegative operator in \cite{bathia2010}, which is used to assess the dimensionality of $X_t(\cdot)$ in equation (\ref{curve.decomp}). 
Third, the proposed method can be applied to both scalar and functional responses with either finite or infinite dimensional functional predictors. To handle a practical scenario where functional predictors are partially observed, we also develop a local linear smoothing approach. Theoretically we establish relevant convergence rates for our proposed estimators under different model settings. In particular, our asymptotic results for partially observed functional predictors reveal interesting phase transition phenomena.
Fourth, empirically we illustrate the superiority of AGMM relative to the potential competitors.

The rest of the paper is organized as follows. In Section~\ref{sec.method}, we present the model for regression with dependent functional errors-in-predictors and develop AGMM fitting procedures for both scalar and functional responses. We also propose the regularized estimator by imposing some form of smoothness into the estimation procedure and discuss the selection of relevant tuning parameters. 
In Section~\ref{sec.thm}, we present convergence results for our proposed estimators for the slope function under different functional scenarios.
In Section~\ref{sec.partial}, we develop a nonparametric smoothing approach for partially observed curve time series and investigate its asymptotic properties.
Section~\ref{sec.emp} illustrates the finite sample performance of AGMM through a series of simulation studies and a public financial dataset. 
All technical proofs are relegated to the Appendix and the Supplementary Material.

\section{Methodology}
\label{sec.method}

\subsection{Model setup}
\label{sec.setup}
In this section, we describe the model setup for the functional linear regression with dependent errors-in-predictors we consider. Let ${\cal L}_2(\cU)$ denote a Hilbert space of square integrable functions defined on $\cU$ equipped with the inner product $ \langle f,g \rangle = \int_{\cU}f(u)g(u)du$ for $f,g \in {\cal L}_2(\cal U).$ Given a scalar response $Y_t,$ a functional predictor $X_t(\cdot)$ in ${\cal L}_2(\cU),$ and, without loss of generality, assuming that $\{Y_t, X_t(\cdot)\}$ have been centered to have mean zero, the classical scalar-on-function linear regression model is of the form 
\begin{equation}
\label{flm.scalar}
Y_t= \int_{\cU} X_t(u) \beta_0(u) du + \varepsilon_t, ~~ t=1, \dots, n,
\end{equation}
where the errors $\varepsilon_t$, independent of $X_{t+k}(\cdot)$ for any integer $k,$ are generated according to a white noise process and $\beta_0(\cdot)$ is the unknown slope function. Generally, $\beta_0$ may not be uniquely determined. We will discuss how to identify $\beta_0$ we wish to estimate later.


We assume that the observed functional predictors $W_1(\cdot), \dots, W_n(\cdot)$ satisfy the error contamination model in equation (\ref{curve.decomp}).
The existence of the unobservable noise term $e_t(\cdot)$ indicates that the curves of interest, $X_t(\cdot),$ are not directly observed. Instead, they are recorded on a grid of points and are contaminated by the error process, $e_t(\cdot),$ without assuming any parametric structure on its covariance function, denoted by $C_e(u,v)=\cov\{e_t(u),e_t(v)\}.$ This model guarantees that all the dynamic elements of $W_t(\cdot)$ are included in the signal term $X_t(\cdot)$ and all the white noise elements are absorbed into the noise term $e_t(\cdot).$ Furthermore, we assume that predictor errors $e_t(\cdot)$ are uncorrelated with both $X_{t+k}(\cdot)$  and $\varepsilon_{t+k},$ for all integer $k.$

Here we turn to discuss the identification of $\beta_0$. Assume that $\big\{\big(Y_t, X_t(\cdot)\big)\big\}$ is strictly stationary and $C_0(u,v)$ is the covariance function of $X_t(\cdot),$ which admits the Karhunen-Lo\`eve expansion,
$X_t(u)=\sum_{j=1}^\infty \xi_{tj} \phi_j(u),$
where $\xi_{tj}=\int_{\cU}X_t(u)\phi_j(u)du$ and $\cov(\xi_{tj},\xi_{tj'})=\lambda_jI(j=j')$ with $I(\cdot)$ denoting the indicator function. Then the eigenpairs $\{\lambda_j, \phi_j(\cdot)\}_{j \geq 1}$ satisfy the eigen-decomposition $\int_{\cU}C_0(u,v)\phi_j(v)dv=\lambda_j\phi_j(u)$ with $\lambda_1 \geq \lambda_2 \geq \cdots \geq 0.$ Define $S_0(u) = E\big\{Y_t X_t(u)\big\},$ $d = {\sup}_{i \ge 1}~\{i: \lambda_i > 0\}$ and assume  $\sum_{j=1}^d \lambda_{j}^{-2} \{\cov(Y_1,\xi_{1j})\}^2 < \infty.$ Obviously $\beta_0$ satisfies the following equation
\begin{equation}
\label{identification_equation}
S_0(u) = \int_{\mathcal{U}} C_0(u,v) \beta(v)dv, u \in \mathcal{U}.
\end{equation}
If the span of eigenfunctions $\{\phi_1, \dots, \phi_d\}$ is dense in the $\mathcal{L}_2$ space, it is clear that $\beta_0$ is the unique solution to (\ref{identification_equation}) and hence can be uniquely identified. In a general scenario, $\beta_0$ can also be well defined.
To make $\beta_0$ identifiable, we consider the following minimization problem
	\begin{equation}
	\label{mini_normal}
	\begin{split}
	 & \min_{ \beta \in \mathcal{L}^2(\mathcal{U})}~ \int_{\cU} \beta^2(u) du, \\
	 & ~s.t.~S_0(u) = \int_{\mathcal{U}} C_0(u,v) \beta(v)dv, ~u \in \mathcal{U}. 
	\end{split}
	\end{equation}
Noting that the solution to (\ref{mini_normal}) exists and is unique, we define the true slope function $\beta_0$ to be this unique minimizer in a closed form of
    $
    \beta_0 = \sum_{j=1}^d \lambda_j^{-1} \cov(Y_1,\xi_{1j}) \phi_j, 
    $ which holds for both $d <\infty$ and $d=\infty.$ 
	See also \cite{cardot2003} and \cite{he2010}. 



\subsection{Main idea}
\label{sec.idea}
In this section, we describe the main idea to facilitate the development of AGMM to estimate $\beta_0(\cdot)$ in (\ref{flm.scalar}). We choose $X_{t+k}(\cdot)$ for $k= 0, 1, \dots,$ as functional instrumental variables, which are assumed to be uncorrelated with the error $\varepsilon_t$ in (\ref{flm.scalar}). Let
\begin{equation}
\label{def.gk}
g_{k}^X(\beta,u) = \cov\big\{Y_t,X_{t+k}(u)\big\}-\int_{\cU}\cov\{X_t(v),X_{t+k}(u)\}\beta(v)dv.
\end{equation}
The population moment conditions, $E\{\varepsilon_tX_{t+k}(u)\}=0$ for any $u \in \cU,$ and equation~(\ref{flm.scalar}) implies that
\begin{equation}
\label{moment.cond}
g_{k}^X(\beta_0,u) \equiv 0 \text{ for any } u\in\cU \text{ and } k=1,\dots.
\end{equation}
In particular, the conventional LS approach is based on  (\ref{moment.cond}) with $k=0.$  However, this approach is inappropriate when $X_t(\cdot)$ are replaced by the surrogates $W_t(\cdot)$ given the fact that $C_W(u,v)=\cov\{W_t(u),W_{t}(v)\}=C_0(u,v) + C_e(u,v),$ and hence the sample version of $C_W(u,v)$ is not a consistent estimator for $C_0(u,v).$ 
See \cite{hall2006} for the identifiability of $C_0(u,v)$ and $C_e(u,v)$ under the assumption that the observed curves $W_1(\cdot),\dots, W_n(\cdot)$ are independent and $e_t(\cdot)$ decays to zero as $n$ goes to infinity. 

To separate $X_t(\cdot)$ from $W_t(\cdot)$ under the serial dependence scenario, we develop a different approach without requiring the ``low noise" condition. For an integer $k\geq 1,$ denote the lag-$k$ autocovariance function of $X_t(\cdot),$ by $C_k(u,v)=\cov\{X_t(u),X_{t+k}(v)\}$, which does not depend on $t.$ Our method is based on the simple fact that $$\cov\{Y_t, W_{t+k}(u)\}=\cov\{Y_t, X_{t+k}(u)\} \text{ and }\cov\{W_t(u),W_{t+k}(v)\}=C_k(u,v) \text{ for any } k \neq 0.$$ 
Then after substituting $X_t(\cdot)$ by $W_t(\cdot)$ in (\ref{def.gk}), we can also represent
$$g_k(\beta,u) = \cov\big\{Y_t,W_{t+k}(u)\big\}-\int_{\cU}\cov\{W_t(v),W_{t+k}(u)\}\beta(v)dv=g_{k}^X(\beta,u),$$ and the moment conditions in (\ref{moment.cond}) become $$g_k(\beta_0,u) \equiv 0 \text{ for any } u \in\cU \text{ and } k=1\dots,L,$$ where $L$ is some prescribed positive integer.

Under the over-identification setting, where the number of moment conditions exceeds the number of parameters, we borrow the idea of {\it generalized methods-of-moments} (GMM) based on minimizing the distance from $g_1(\beta,\cdot), \dots, g_L(\beta,\cdot)$ to zero. This distance is defined by the quadratic form of
$$Q(\beta)=\sum_{k=1}^L\sum_{l=1}^L\int_{\cU}\int_{\cU}g_k(\beta,u)\Omega_{k,l}(u,v)g_l(\beta,v)dudv,$$ where $\bOmega(u,v)=\{\Omega_{k,l}(u,v)\}_{1 \leq k,l \leq L}$ is an $L$ by $L$ weight matrix whose $(k,l)$-th element is $\Omega_{k,l}(u,v).$ A suitable choice of $\bOmega(u,v)$ must satisfy the properties of symmetry and positive-definiteness \cite[]{guh2013}, which are, to be specific, (i) $\Omega_{kl}(u,v)=\Omega_{lk}(v,u)$ for each $k,l=1,\dots,L$ and $(u,v)\in \cU^2;$ (ii) for any finite collection of time points $u_1,\dots, u_T,$ $\sum_{t=1}^T\sum_{t'=1}^T \ba(u_t)^{\T}\bOmega(u_t,u_{t'})\ba(u_{t'})$ must be positive for any $\ba(\cdot)=\big(a_1(\cdot), \dots, a_L(\cdot)\big)^{\T}.$
In general, one can choose the optimal weight matrix $\bOmega$ and implement a two-step GMM. However, this would give a very slight improvement in our simulations. To simplify our derivation and accelerate the computation, we choose the identity weight matrix as $\Omega_{k,l}(u,v)=I(k=l)I(u=v)$ and then minimize the resulting distance of 
$$  Q(\beta)=\sum_{k=1}^{L}\int_{\cU}g_k(\beta,u)^2du,$$
over $\beta(\cdot) \in {\cal L}_2(\cU).$ The minimizer of $Q(\beta)$, $\beta_0(\cdot)$, can be achieved by solving $\partial Q(\beta)/\partial \beta=0,$ i.e. for any $u\in\cU,$
\begin{equation}
\label{gmm.sol}
\sum_{k=1}^L \left[ \int_{\cU}C_k(u,z)\cov\big\{Y_t,W_{t+k}(z)\big\}dz-\int_{\cU}\Big\{\int_{\cU}C_k(u,z)C_k(v,z)dz\Big\}\beta(v)dv\right] = 0.    
\end{equation}

To ease our presentation, we define 
\begin{equation}
\label{def.R}
R(u)=\sum_{k=1}^L \int_{\cU}C_k(u,z)\cov\{Y_t,W_{t+k}(z)\}dz
\end{equation}
and
\begin{equation}
\label{def.K}
K(u,v)=\sum_{k=1}^L \int_{\cU}C_k(u,z)C_k(v,z)dz.
\end{equation}
Note that $K$ can be viewed as the kernel of a linear operator acting on ${\cal L}_2(\cU),$ i.e. for any $f\in {\cal L}_2(\cU),$ $K$ maps $f(u)$ to $\widetilde f(u)\equiv\int_{\cU}K(u,v)f(v)dv.$ For notational economy, we will use $K$ to denote both the kernel and the operator. Indeed, the nonnegative definite operator $K$ was proposed in \cite{bathia2010} to identify the dimensionality of $X_t(\cdot)$ based on $W_t(\cdot)$ in (\ref{curve.decomp}).
Substituting the relevant terms in (\ref{gmm.sol}), $\beta_0(\cdot)$ satisfies the following equation
\begin{equation}
\label{gmm.sol.normal}
R(u)=\int_{\cU}K(u,v)\beta(v)dv \text{ for any } u \in \cU.
\end{equation} See also functional extension of the least squares type of normal equation in (\ref{identification_equation}). 

Provided that $X_t(\cdot)$ is $d$-dimensional, it follows from Proposition~1 of \cite{bathia2010} that, under regularity conditions, $K$ has the spectral decomposition, $K(u,v)=\sum_{j=1}^d \theta_j \psi_j(u) \psi_j(v),$ with $d$ nonzero eigenvalues $\theta_1 \geq \theta_2 \geq \dots\geq \theta_d$ and $\overline{\mbox{span}}\{\psi_1,\ldots,\psi_d\}$ is the linear space spanned by the $d$ eigenfunctions  $\{\phi_1, \dots, \phi_d\}.$ This assertion still holds even for $d=\infty.$ 

Denote the null space of $K$ and its orthogonal complement by $\ker(K)=\{x \in {\cal L}_2(\cU): Kx=0\}$  and 
${\ker(K)}^{\perp}=\{x \in {\cal L}_2(\cU): \langle x,y \rangle =0, \forall y \in \ker(K)\},$ respectively. The inverse operator $K^{-1}$ corresponds to the inverse of the restricted operator $\breve K = K\big| \ker(K)^{\perp}$, which restricts the domain of $K$ to $\ker(K)^{\perp}.$  See Section~3.5 of \cite{Bhsing2015} for details. When $ d < \infty,$ $\beta_0(\cdot)$  is indeed the unique solution to (\ref{gmm.sol.normal}) in $\ker(K)^{\perp}$ in the form of
\begin{equation}
\label{beta.sol.finite}
\beta_0(u) = \int_{\cU}K^{-1}(u,v)R(v)dv=\sum_{j=1}^{d} \theta_j^{-1} \langle \psi_j,R\rangle \psi_j(u).
\end{equation}
Provided $K$ is a bounded operator when $d=\infty$, $K^{-1}$ becomes an unbounded operator, which means it is discontinuous and cannot be estimated in a meaningful way. However, $K^{-1}$ is usually associated with another function/operator, the composite function/operator can be reasonably assumed to be bounded, e.g. the regression operator \cite[]{li2018}. If we further assume that the composite function $\int_{\cU}K^{-1}(u,v)R(v)dv$ is bounded, or equivalently $\sum_{j=1}^{\infty}\theta_j^{-2} \langle \psi_j, R \rangle^2 < \infty,$ $\beta_0(\cdot)$  is still the unique solution to (\ref{gmm.sol.normal}) in $\ker(K)^{\perp}$ and is of the form
\begin{equation}
\label{beta.sol.inf}
\beta_0(u)=\int_{\cU}K^{-1}(u,v)R(v)dv=\sum_{j=1}^{\infty}
\theta_j^{-1} \langle \psi_j, R \rangle \psi_j(u).
\end{equation}
Both (\ref{beta.sol.finite}) and (\ref{beta.sol.inf}) motivate us to develop the estimation procedure for $\beta_0$ in Section~\ref{sec.est}.
\subsection{Estimation procedure}
\label{sec.est}
In this section, we present the AGMM estimator for $\beta_0(\cdot)$ based on the main idea described in Section~\ref{sec.idea}.

We first provide the estimates of $C_k(u,v)$ and $\cov\{Y_t,W_{t+k}(u)\}$ for $k=1, \dots, L,$ i.e.
\begin{equation}
\label{est.C}
\widehat C_k(u,v) =  \frac{1}{n-L}\sum_{t=1}^{n-L} W_t(u)W_{t+k}(v)
\quad \text{and} \quad
\widehat{\cov}\{Y_t,W_{t+k}(u)\} =  \frac{1}{n-L}\sum_{t=1}^{n-L} Y_tW_{t+k}(u).
\end{equation}  
Combing (\ref{def.R}), (\ref{def.K}) and (\ref{est.C}) gives the the natural estimators for $K(u,v)$ and $R(u)$ as
\begin{equation} 
\label{est.K}
\widehat K(u,v) =\sum_{k=1}^L \int_{\cU}\widehat C_k(u,z)\widehat C_k(v,z)dz = \frac{1}{(n-L)^2}\sum_{k=1}^L\sum_{t=1}^{n-L}\sum_{s=1}^{n-L} W_t(u)W_s(v) \langle W_{t+k},W_{s+k}\rangle
\end{equation}
and
\begin{equation}
\label{est.R}
\widehat R(u) = \sum_{k=1}^L \int_{\cU}\widehat C_k(u,z)\widehat{\cov}\{Y_t,W_{t+k}(z)\}dz=\frac{1}{(n-L)^2}\sum_{k=1}^L\sum_{t=1}^{n-L}\sum_{s=1}^{n-L} W_t(u)Y_s \langle W_{t+k},W_{s+k} \rangle,
\end{equation}
respectively.
Note we choose a fixed integer $L>1,$ as $K$ pulls together the information at different lags, while $L=1$ may lead to spurious estimation results. See Section~\ref{sec.tune} for the discussion on the selection of $L.$

We next perform an eigenanalysis on $\widehat K$ and thus obtain the estimated eigenpairs $\{\widehat\theta_j, \widehat\psi_j(\cdot)\}$ for $j=1,2,\dots.$ When the number of functional observations $n$ is large, the accumulated errors in (\ref{est.K}), (\ref{est.R}) and the eigenanalysis on $\widehat K$ are relatively small, thus resulting in smooth estimates of $\psi_j(\cdot)$ and $\beta_0(\cdot).$ We refer to this implementation of our method as Base AGMM for the remainder of the paper. However, in the setting without a sufficiently large $n$ this version of AGMM suffers from a potential under-smoothing problem that the resulting estimate of $\beta_0(\cdot)$ wiggles quite a bit. To overcome this disadvantage, we can impose some level of smoothing in the eigenanalysis through the basis expansion approach, which converts the continuous functional eigenanalysis problem for $\widehat K$  to an approximately equivalent matrix eigenanalysis task. We explore this {\it basis expansion based AGMM}, simply referred to as AGMM from here on. To be specific, let $\bB(u)$ be the $J$-dimensional orthnormal basis function, i.e. $\int_{\cU} \bB(u) \bB^{\T}(u)du = \bI_{J},$ such that for each $j =1,\ldots,J$, $\psi_j(\cdot)$ can be well approximated by $\bdelta_j^{\T}\bB(\cdot),$ where $\bdelta_j$ is the basis coefficients vector. Let 
$$\widehat{\bK} = \int_{\cU} \int_{\cU} \bB(u) \bB^{\T}(v) \widehat {K}(u,v)du dv.$$
Performing an eigen-decomposision on $\widehat{\bK}$ leads to the estimated eigenpairs $\{(\widehat \theta_j, \widehat{\bdelta}_j)\}_{j=1}^J$. Then the $j$-th estimated principal component function is given by $\widehat{\psi}_j(\cdot) = \widehat{\bdelta}_j^{\T} \bB(\cdot).$ See Section~\ref{sec.tune} for the selection of $J$. A similar basis expansion technique can be applied to produce a smooth estimate $\widehat R(\cdot).$ Note that $\widehat \bK, \widehat \theta_j, \widehat\psi_j, j=1, \dots,d,$ all depend on $J,$ but for simplicity of notation, we will omit the corresponding superscripts where the context is clear.

Finally, we substitute the relevant terms in (\ref{beta.sol.finite}) and (\ref{beta.sol.inf}) by their estimated values. We discuss two situations corresponding to $d<\infty$ and $d=\infty$ as follows. 
(i) When $X_t(\cdot)$ is $d$-dimensional ($d<\infty$), we need to select the estimate $\widehat d$ of $d$ in the sense that $\widehat\theta_1, \dots, \widehat\theta_{\widehat d}$ are large eigenvalues of $\widehat K$ and $\widehat \theta_{\widehat d +1}$ drops dramatically. The estimate $\widehat\beta$ of $\beta_0$ is then given by
\begin{equation}
\label{beta.est.finite}
\widehat \beta(u)=\sum_{j=1}^{\widehat d} \widehat \theta_j^{-1} \langle \widehat \psi_j, \widehat R \rangle  \widehat \psi_j(u).
\end{equation} 
(ii) When $X_t(\cdot)$ is an infinite dimensional functional object,  we take the standard truncation approach by using the leading $M$ eigenpairs of $\widehat K$ to approximate $\beta_0$ in (\ref{beta.sol.inf}). Specifically, we obtain the estimated slope function as
\begin{equation}
\label{beta.est.inf}
\widehat \beta(u)=\sum_{j=1}^{ M} \widehat \theta_j^{-1} \langle \widehat \psi_j, \widehat R \rangle  \widehat \psi_j(u).
\end{equation} 

Section~\ref{sec.tune} presents details to select $\widehat d$ and $M.$ However, when $d=\infty,$ the empirical performance of $\widehat{\beta}(\cdot)$ may be sensitive to the selected value of $M.$ To improve the numerical stability, we suggest an alternative ridge-type method to estimate $\beta_0.$ Specifically, we propose
\begin{equation}
\label{ridge.est}
\widehat \beta_{\text{ridge}}(u)=\sum_{j=1}^{\widebar M} (\widehat \theta_j + \rho_{n})^{-1} \langle \widehat \psi_j, \widehat R \rangle  \widehat \psi_j(u),
\end{equation} 
where $\widebar M$ is chosen to be reasonably larger than $M$ and $\rho_n \geq 0$ is a ridge parameter. See also \cite{hall2007} for the ridge-type estimator in classical functional linear regression. 

\subsection{Generalization to functional response}
\label{sec.func}
In this section, we consider the case when the response is also functional. Given a functional response $Y_t(\cdot)$ and a functional predictor $X_t(\cdot)$, both of which are in ${\cal L}_2(\cU)$ and have mean zero, the function-on-function linear regression takes the form of  
\begin{equation}
\label{flm.func}
Y_t(u)= \int_{\cU} X_t(v) \gamma_0(u,v) dv + \varepsilon_t(u), \ u \in \cU, \ t=1,\dots, n,
\end{equation}
where $\gamma_0(u,v)$ is the slope function of interest and $\varepsilon_t(\cdot)$, independent of $X_{t+k}(\cdot)$ for any integer $k$, are random elements in the underlying separable Hilbert space. We still observe the erroneous version $W_t(\cdot)$ rather than the signal $X_t(\cdot)$ itself in equation (\ref{curve.decomp}). 

To estimate the slope function in (\ref{flm.func}), we develop an AGMM approach analogous to that for the scalar case in Section~\ref{sec.method} by solving the normal equation of
\begin{equation}
\label{gmm.sol.normal.fun}
H(u,v)=\int_{\cU}K(u,w)\gamma(w,v)dw  \text{ for any } v \in \cU,
\end{equation}
where $H(u,v)=\sum_{k=1}^L \int_{\cU} C_k(u,z)\cov\{Y_t(v),W_{t+k}(z)\}dz$ with its natural estimator
\begin{equation}
\label{est.H}
\widehat H(u,v) = \frac{1}{(n-L)^2}\sum_{k=1}^L\sum_{t=1}^{n-L}\sum_{s=1}^{n-L} W_t(u)Y_s(v)\langle W_{t+k},W_{s+k}\rangle.
\end{equation}

Accordingly, we can provide the estimate $\widehat\gamma$ of $\gamma_0$ under two functional scenarios including $d<\infty$ and $d=\infty$. (i) When $d<\infty,$ $\gamma_0(u,v)$ is the unique solution of (\ref{gmm.sol.normal.fun}) in $\ker(K)^{\perp}$ and can be represented as
\begin{equation}
\label{beta.sol.fun}
\gamma_0(u,v)=\int_{\cU}K^{-1}(u,w)H(w,v)dw=\sum_{j=1}^{d}\theta_j^{-1} \langle \psi_j,H(\cdot,v)\rangle \psi_j(u).
\end{equation}
The estimate of $\gamma_0(u,v)$ is then given by
\begin{equation}
\label{beta.est.finite.fun}
\widehat \gamma(u,v)=\sum_{j=1}^{\widehat d} \widehat\theta_j^{-1} \widehat \psi_j(u)\langle \widehat\psi_j, \widehat H(\cdot,v)\rangle.
\end{equation}
(ii) Under the infinite dimensional setting ($d=\infty,$ if we assume the boundedness of the composite function $\int_{\cU}K^{-1}(u,w)H(w,v)dw$ in the $L_2$ sense, 
the solution to (\ref{gmm.sol.normal.fun}) uniquely exists. Approximating the infinite dimensional $\gamma_0(u,v)$ in (\ref{beta.sol.fun}) by the first $M$ components and substituting the relevant terms by their estimated values, we can obtain 
\begin{equation}
\label{beta.est.inf.fun}
\widehat \gamma(u,v)=\sum_{j=1}^{M}\widehat\theta_j^{-1} \widehat \psi_j(u) \langle \widehat\psi_j, \widehat H(\cdot,v)\rangle .
\end{equation}

\subsection{Selection of tuning parameters}
\label{sec.tune}
Implementing AGMM requires choosing $L$ (selected lag length in (\ref{gmm.sol})), $M$ (truncated dimension in (\ref{beta.est.inf}) when $d=\infty$), $\widehat d$ (number of identified nonzero eigenvalues of $\widehat K$ when $d<\infty$) and $J$ (dimension of the basis function $\bB(u)$). First, we tend to select a small value of $L,$ as the strongest autocorrelations usually appear at the small time lags and adding more terms will make $\widehat K$ less accurate. Our simulated results suggest that the proposed estimators are not sensitive to the choice of $L,$ therefore we set $L=5$ in our empirical studies. See also \cite{bathia2010} and \cite{lam2011} for relevant discussions. 

Second, to select $M$ when  $d=\infty,$ the typical approach is to find the largest $M$ eigenvalues of $\widehat K$ such that the corresponding cumulative percentage of variation exceeds the pre-specified threshold value, e.g. 90\% or 95\%. Other available methods 
include the bootstrap test \cite[]{bathia2010} and the eigen-ratio-based estimator \cite[]{lam2011}. Third, to determine $\widehat d$ when $d < \infty,$ we take the bootstrap approach proposed in \cite{bathia2010}. Our task is to test the null hypothesis $H_0: \theta_{d+1}=0$. We reject $H_0$ if $\widehat\theta_{d+1}>c_{\alpha},$ where $c_{\alpha}$ is the critical value corresponding to the significant level $\alpha \in (0,1).$ We summarize the bootstrap procedure as follows.
\begin{enumerate}
	\item Define $\widehat W_t(\cdot)=\sum_{j=1}^{\widehat d}\widehat \eta_{tj}\widehat\psi_{j}(\cdot),$ where $\widehat \eta_{tj}=\int_{\cU}W_{t}(u)\widehat{\psi}_j(u)du$ for $j=1, \dots, \widehat d.$ Let $\widehat e_t(\cdot)=W_t(\cdot)-\widehat W_t(\cdot).$
	\item Generate a bootstrap sample using $W_t^*(\cdot)=\widehat W_t(\cdot)+ e_t^*(\cdot),$ where $e_t^*$ are drawn with replacement from $\{\widehat e_1, \dots, \widehat e_n\}.$
	\item 
	In an analogy to $\widehat K$ defined in (\ref{est.K}), form an estimator $\widehat K^*$ by replacing $\{W_t\}$ with $\{W_t^*\}.$ Then calculate the $(d+1)$-th largest eigenvalue $\theta_{d+1}^*$ of $\widehat K^*.$
\end{enumerate}
We repeat Steps~2 and 3 above $B$-times and reject $H_0$ if the event of $\{\widehat \theta_{d+1}>\theta_{d+1}^*\}$ occurs more than $[(1-\alpha)B]$ times. 
Starting with $\widehat d=1,$ we sequentially test $\theta_{{\widehat d+1}}=0$ and increase $\widehat d$ by one until the resulting null hypothesis fails to be rejected.

Fourth, to select $J,$ we propose the following $G$-fold cross-validation (CV) approach.
\begin{enumerate}
	\item Sequentially divide the set $\{1, \dots,n\}$ into $G$ blockwise groups, $\cD_1, \dots, \cD_G,$ of approximately equal size. 
	
	\item Treat the $g$-th group as a validation set. Implement the regularized eigenanalysis in Section~\ref{sec.est} on the remaining $G-1$ groups, compute $\widehat \bK^{(-g)}$ and let 
	$\widehat\bdelta_1^{(-g)}, \dots, \widehat\bdelta_d^{(-g)}$ be the top $d$ eigenvectors of $\widehat \bK^{(-g)}.$
	
	\item Compute $\widehat K^{(g)}(u,v)$ and $\widehat \bK^{(g)}$ based on the validation set. 
	Let $\widehat\theta_l^{(g)}=(\widehat\bdelta_l^{(-g)})^{\T}\widehat \bK^{(g)}\widehat\bdelta_l^{(-g)}$ for $l=1, \dots, d.$
\end{enumerate}
We repeat Steps~2 and 3 above $G$ times and choose $J$ as the value that minimize the following mean CV error
$$\text{CV}(J)=\frac{1}{G}\sum_{g=1}^G \int_{\cU}\int_{\cU} \Big\{\widehat K^{(g)}(u,v)-\sum_{j=1}^d\widehat \theta_j^{(g)}(\widehat\bdelta_j^{(-g)})^{\T}\bB(u)\bB(v)^{\T}\widehat\bdelta_j^{(-g)} \Big\}^2dudv.$$ Given the time break on the training observations, the autocovariance assumption is jeopardized by $L=5$ misutilized lagged terms. However, this effect on $\widehat K$ is negligible especially when $n$ is sufficiently large, hence our proposed CV approach can still be practically applied. See also \cite{bergmeir2018} for various CV methods for time dependent data. 

\section{Theoretical properties}
\label{sec.thm}
In this section, we investigate the theoretical properties of our proposed estimators for both scalar-on-function and function-on-function linear regressions. 

To present the asymptotic results, we need the following regularity conditions.
\begin{condition}
	\label{cond_mix}
	$\{W_{t}(\cdot), t=1,2,\dots\}$ is strictly stationary curve time series. Define the $\psi$-mixing with the mixing coefficients $$\psi(l) = \underset{A \in \cF_{-\infty}^0, B \in \cF_{l}^{\infty}, P(A)P(B)>0}{\sup} |1-P(B|A)/P(B)|, ~~l=1, 2, \dots,$$
	where $\cF_{i}^j$ denotes the $\sigma$-algebra generated by $\{W_t(\cdot), i \leq t \leq j\}.$ Moreover, it holds that $\sum_{l=1}^{\infty} l \psi^{1/2}(l) < \infty.$
\end{condition}
\begin{condition}
	\label{cond_moment_scalar}
	$E(\|W_t\|^4) < \infty$ and $E(\varepsilon_t^2) < \infty.$ 
\end{condition}

The presentation of the $\psi$-mixing condition in Condition~\ref{cond_mix} is mainly for technical convenience. See Section~2.4 of \cite{Bbosq2000} on the mixing properties of curve time series. Condition~\ref{cond_moment_scalar} is the standard moment assumption in functional regression literature \cite[]{hall2007,chak2017}. 

\begin{condition}
	\label{cond_eigen}
	(i) When $d$ is fixed, $\theta_1> \cdots > \theta_d>0=\theta_{d+1};$ 
	(ii) When $d=\infty,$ $\theta_1> \theta_2 > \cdots >0,$ and there exist some positive constants $c$ and $\alpha>1$ such that $ \theta_j-\theta_{j+1} \geq cj^{-\alpha-1}$ for $j \ge 1$;
	(iii) $\overline{\mbox{span}}\{\phi_1,\ldots,\phi_d\} =  \overline{\mbox{span}}\{\psi_1, \dots, \psi_d\}.$ 
\end{condition}

\begin{condition}
	\label{cond_bias_scalar} 
	When $d=\infty,$ $\beta_0(u) = \sum_{j = 1}^\infty b_j \psi_j(u)$ and there exist some positive constants $\tau \ge \alpha+1/2$ and $C$ such that 
	$|b_j| \le C j^{-\tau}$ for $j \geq 1.$
\end{condition}

Condition~\ref{cond_eigen} restricts the eigen-structure of $K$ and assumes that all the nonzero eigenvalues of $K$ are distinct from each other. When $d=\infty,$ Condition~\ref{cond_eigen} (ii) prevents gaps between adjacent eigenvalues from being too small. The parameter $\alpha$ determines the tightness of eigen-gaps with larger values of $\alpha$ yielding tighter gaps. This condition also indicates that $\theta_j \geq c\alpha^{-1}j^{-\alpha}$ as $\theta_j=\sum_{k=j}^{\infty}(\theta_k-\theta_{k+1})\geq c\sum_{k=j}^{\infty} k^{-\alpha-1},$ and can be used to derive the convergence rates of estimated eigenfunctions.
See also \cite{hall2007} and \cite{qiao2018}. 
Condition~\ref{cond_bias_scalar} restricts $\beta_0$ based on its expansion using eigenfunctions of $K.$ The parameter $\tau$ determines the decay rate of slope basis coefficients, $\{b_j\}_{j=1}^{\infty}$. The assumption $\tau \ge \alpha+1/2$ can be interpreted as requiring $\beta_0$ be sufficiently smooth relative to $K$, the smoothness of which can be implied by $\theta_j \geq c\alpha^{-1}j^{-\alpha}$ from Condition~\ref{cond_eigen}~(ii). See \cite{hall2007} for an analogous condition in functional linear regression.

Before presenting Theorem~\ref{thm_est_scalar} 
for the asymptotic analysis of the scalar-on-function linear regression,
we first solidify some notation. For any univariate function $f,$ define $\|f\|=\sqrt{\langle f,f\rangle}.$ We denote by $\|A\|_{\cS}$ the Hilbert-Schmidt norm for any bivariate function $A.$ The notation $a_n \asymp b_n$ for positive $a_n$ and $b_n$ means that the ratio $a_n/b_n$ is bounded away from zero and infinity. To obtain $\widehat\beta$ in (\ref{beta.est.finite}) when $d<\infty,$ we use the consistent estimator for $d$ defined as $\widehat d = \#\{j:\widehat \theta_j \geq \epsilon_n\},$ where $\epsilon_n$ satisfies the condition in Theorem~\ref{thm_est_scalar}~(i) below. Then by Theorem~3 of \cite{bathia2010}, $\widehat d$ converges in probability to $d$ as $n \rightarrow \infty.$ 

\begin{theorem}
	\label{thm_est_scalar}
	Suppose that Conditions~\ref{cond_mix}--\ref{cond_bias_scalar} hold.  The following assertions hold as $n\rightarrow \infty:$\\
	(i) Let $\epsilon_n \rightarrow 0$ and $\epsilon_n^2n \rightarrow \infty$ as $n \rightarrow \infty.$ When $d$ is fixed, then $$\|\widehat \beta- \beta_0\|=O_P\big(n^{-1/2}\big).$$
	(ii) When $d = \infty,$ if we further assume that $M\asymp n^{1/(2\alpha + 2\tau)},$ then 
	$$
	\|\widehat \beta- \beta_0\|^2 = O_P\big(M^{2\alpha +1}n^{-1} + M^{-2\tau +1}\big) = O_P\big( n^{- \frac{2 \tau - 1}{2\alpha+2\tau}}\big).
	$$
\end{theorem}

{\bf Remarks.} 
(a) When $d$ is fixed, the standard parametric root-$n$ rate is achieved. 

(b) When $d=\infty,$ the convergence rate is governed by two sets of parameters (1) dimensionality parameter, sample size ($n$); (2) internal parameters, truncated dimension of the curve time series ($M$), decay rate of the lower bounds for eigenvalues ($\alpha$), decay rate of the upper bounds for slope basis coefficients ($\tau$). It is easy to see that larger
values of $\alpha$ (tighter eigen-gaps) yield a slower convergence rate, while increasing $\tau$ enhances the smoothness of $\beta_0(\cdot),$ thus resulting in a faster rate. The convergence rate consists of two terms, which reflects our familiar variance-bias tradeoff as commonly considered in nonparametric statistics. In particular, the bias is bounded by $O(M^{-\tau +1/2})$ and the variance is of the order $O_P(M^{2\alpha+1}n^{-1}).$ To balance both terms, we choose the truncated dimension, $M \asymp n^{1/(2\alpha+2\tau)},$ while the optimal convergence rate then becomes $O_P\{n^{-(2 \tau - 1)/(2\alpha+2\tau)}\}.$  It is also worth noting that this rate is slightly slower than the minimax rate $O_P\{n^{-(2 \tau - 1)/(\alpha+2\tau)}\}$ in \cite{hall2007}, which considers independent observations of the functional predictor without any error contamination. In fact, we tackle a more difficult functional linear regression scenario, 
where extra complications come from the serial dependence and functional error contamination. From a theoretical perspective, whether the rate in part~(ii) is optimal in the minimax sense is still of interest and requires further investigation.



Before presenting the asymptotic results for the function-on-function linear regression, we list Conditions~\ref{cond_moment_func} and \ref{cond_bias_func} below, which are substitutes of Conditions~\ref{cond_moment_scalar} and \ref{cond_bias_scalar}, respectively, in the functional response case.

\begin{condition}
	\label{cond_moment_func}
	$E(\|W_t\|^4) < \infty$ and $E(\|\varepsilon_t\|^2) < \infty.$
\end{condition}
\begin{condition}
	\label{cond_bias_func} 
	When $d=\infty,$ $\gamma_0(u,v) = \sum_{j=1}^\infty\sum_{\ell=1}^\infty b_{j\ell} \psi_j(u) \psi_{\ell}(v)$ and there exist some positive constants $\tau \geq \alpha+1/2$ and $C$ such that 
	$
	|b_{j\ell}| \le C (j+\ell)^{-\tau - 1/2}
	$ for $j, \ell \geq 1.$
\end{condition}
\begin{theorem}
	\label{thm_est_func}
	Suppose that Conditions~\ref{cond_mix}, \ref{cond_eigen}, \ref{cond_moment_func} and \ref{cond_bias_func} hold.  The following assertions hold as $n\rightarrow \infty:$\\
	(i) Let $\epsilon_n \rightarrow 0$ and $\epsilon_n^2n \rightarrow \infty$ as $n \rightarrow \infty.$ When $d$ is fixed, then
	$$
	\|\widehat \gamma- \gamma_0\|_{\cS}=O_P(n^{-1/2}).
	$$
	(ii) When $d = \infty$, if we further assume that $M\asymp n^{1/(2\alpha + 2\tau)},$ then 
	$$
	\|\widehat \gamma- \gamma_0\|_{\cS}^2=O_P\big(M^{2\alpha +1}n^{-1} + M^{-2\tau +1}\big) =O_P\big(n^{-\frac{2 \tau - 1}{2 \alpha +2 \tau}}\big).
	$$
\end{theorem}

\section{Partially observed functional predictor}
\label{sec.partial}

In this section, we consider a practical scenario where each $W_t(\cdot)$ is partially observed at random time points, $U_{t1}, \dots, U_{t m_t} \in \cU=[0,1],$ where for dense measurement designs all $m_t$'s are larger than some order of $n$, and for sparse designs all $m_t$'s are bounded \cite[]{zhang2016,qiao2020}.
Let $Z_{ti}$ represent the observed value of $W_{t}(U_{t_i})$ satisfying
\begin{equation}\label{curve.patial}
    Z_{ti} = W_t(U_{ti}) + \eta_{ti}, ~~i = 1,\ldots, m_t, 
\end{equation}
where $\eta_{ti}$'s are i.i.d. random errors with finite variance, independent of $W_t(\cdot)$. 

Let $K(\cdot)$ be an univariate kernel function.
We apply a local linear surface smoother to estimate the lag-$k$ autocovariance function $C_{k}(u,v)$ for $k=1, \dots, L$ by minimizing 
\begin{equation}
\label{smooth.autocov}
\sum_{t = 1 }^{n - L} \sum_{i=1}^{m_t}\sum_{j = 1}^{m_{t+k}} 
 \left\{ Z_{ti} Z_{(t+k)j} - a_{0}^{(k)} - a_1^{(k)}(U_{ti} -u) - a_2^{(k)}(U_{(t+k)j} - v)\right\}^2 K_{k,i,j,t,h}(u,v) 
\end{equation}
with respect to $(a_0^{(k)}, a_1^{(k)}, a_2^{(k)}),$ where $K_{k,i,j,t,h}(u,v) = K\left(\frac{U_{ti} - u}{h_C}\right)
K\left(\frac{U_{(t+k)j} - v}{h_C}\right)$ with a bandwidth $h_C>0.$
Let the minimizer of (\ref{smooth.autocov}) be $(\widehat a_0^{(k)}, \widehat a_1^{(k)}, \widehat a_2^{(k)})$ and the resulting lag-$k$ autocovariance estimator is $\widetilde{C}_k(u,v) = \widehat a_0^{(k)}.$ Similarly, we implement a local linear smoothing approach to estimate $S_k(u) = \cov(Y_t, W_{t+k}(u))$ for $k=1, \dots, L$ by minimizing
\begin{equation}
\label{smooth.response}
\sum_{t = 1 }^{n - L} \sum_{i=1}^{m_t} \left\{ Y_t Z_{(t+k)i}  - b_0^{(k)} - b_1^{(k)} (U_{(t+k)i} - u) \right\}^2 K\left(\frac{U_{ti} - u}{h_S}\right)
\end{equation}
with respect to $(b_0^{(k)}, b_1^{(k)})$ with a bandwidth $h_S>0.$ Then we obtain the estimate $\widetilde{S}_k(u) = \widehat{b}_0^{(k)}.$ We also develop a basis expansion approach \cite[]{radchenko2015} to estimate $C_k$ and $S_k,$ where details can be found in Section~\ref{sec.basis.app} 
of the Supplementary Material. 

Let $\widetilde{K}(u,v) = \sum_{k=1}^L \int_{\cU}\widetilde{C}_k(u,z) \widetilde{C}_{k}(v,z)dz$ with estimated eigenpairs $(\widetilde \theta_j, \widetilde \psi_j)_{j\geq 1}$ and 
 $\widetilde{R}(u) = \sum_{k=1}^L \int_{\cU} \widetilde{C}_k(u,z) \widetilde{S}_k(z)dz$. In analogy to (\ref{beta.est.finite}) and (\ref{beta.est.inf}), we obtain the corresponding estimates $\widetilde{\beta}$ of $\beta_0$ by replacing $(\widehat\theta_j,\widehat\psi_j)_{j \geq 1}$ and $\widehat R$ with $(\widetilde\theta_j,\widetilde\psi_j)_{j \geq 1}$ and $\widetilde R,$ respectively.  Before presenting the main asymptotic results, we impose the following regularity conditions.

\begin{condition}
\label{cond_discrete_mixing}
   (i) The errors $\{\eta_{ti}\}$ are $i.i.d.$ mean zero random variables with $E|\eta_{ti}|^{2s} < \infty$ for some $s > 2;$
  (ii) $\{W_t(\cdot), t = 1,2,\cdots\}$ is strictly stationary with $\psi$-mixing coefficients $\psi(l)$ satisfying
$
\psi(l) \lesssim  l^{-\lambda}
$
with $\lambda > \frac{3s- 2}{s-2}$
and $\sup_{u\in [0,1]}E|W_t(u)|^{2s} < \infty.$ 
\end{condition}

\begin{condition}
\label{cond_kernel}
$K(\cdot)$ is a symmetric probability density function on $[-1, 1]$ 
and is Lipschitz continuous.
\end{condition}
\begin{condition}
\label{cond_time}
$\{U_{ti}, i =1,\ldots,m_t\}$ are i.i.d. copies of a random variable $U$ defined on $[0,1]$ and the density $f(\cdot)$ of $U$ is twice continuously differentiable and is bounded from below and above over $[0,1].$
\end{condition}

\begin{condition}
    \label{cond_independence}
    $\{W_t\}$ are independent of $\{U_{ti}\}$ and $\{\eta_{ti}\}$ are independent of $\{U_{ti}\}, \{W_t\}.$
\end{condition}
\begin{condition}
    \label{cond_2der}
    (i) $\partial^2C_k(u,v)/\partial u^2,\partial^2C_k(u,v)/\partial u\partial v$ and $\partial^2C_k(u,v)/\partial v^2$ for $k \geq 1$ are uniformly continuous and bounded on $[0,1]^2;$ 
    (ii) $\partial^2S_k(u)/\partial u^2$ for $k \geq 1$ are uniformly continuous and bounded on $[0,1].$
\end{condition}
\begin{condition}
	\label{cond_observation_times}
The number $m_t$ of measurement locations in time $t$ are independent random variables with distribution $m_t\rho_n^{-1} \sim \widecheck{m},$ where $\widecheck{m} \in \{1,\ldots,\widebar{m}\}$ for some bounded $\widebar{m}$ such that $P(\widecheck{m} > 1) > 0.$
\end{condition}

\begin{condition}
	\label{cond_bandwidths}
The bandwidth parameters $h_C$ and $h_S$ satisfy
$$
h_C \to 0, ~h_S \to 0, ~\frac{\log (n \rho_n^2)}{(n \rho_n^2)^{\theta_C} h_C^2}  \to 0~\mbox{and}~ \frac{\log (n \rho_n)}{(n \rho_n)^{\theta_S} h_S} \to 0,
$$
with
$$
\theta_{C} = \frac{\beta - 2 - (1+\beta)/(s-1)}{\beta + 2 - (1+ \beta)/(s-1)}, ~~
\theta_{S} = \frac{\beta - 3 - (1+\beta)/(s-1)}{\beta + 1 - (1+ \beta)/(s-1)}.
$$
\end{condition}


Conditions~\ref{cond_discrete_mixing}--\ref{cond_bandwidths} are standard in local linear smoothing when the serial dependence exists \cite[]{hansen2008,rubin2020}. In Condition~\ref{cond_observation_times}, we treat the number $m_t$ of measurement locations as random variables, but possibly diverges with $n$ at the order of $\rho_n.$ When $\rho_n$ is bounded, it corresponds to the sparse case in \cite{rubin2020}. 

We present the convergence rates of $\widetilde{C}_k, \widetilde{S}_k$ for $k \geq 1$  and $\widetilde \beta_0$ in the following Theorems~\ref{thm_covariance_discrete} and \ref{thm_beta_discrete}, respectively.
\begin{theorem}
	\label{thm_covariance_discrete}
	Suppose that Conditions~\ref{cond_discrete_mixing}--\ref{cond_bandwidths} hold. As $n \to \infty,$ we have
$$
\|\widetilde{C}_k - C_k\|_{\cS} = O_P\left( \delta_{n1}\right) ~\mbox{and}~~
\|\widetilde{S}_k - S_k\| = O_P\left( \delta_{n2}\right) \text{ for } k \geq 1, 
$$
where $$
\delta_{n1} = \frac{1}{\sqrt{n \rho_n^2 h_C^2}} +\frac{1}{\sqrt{n}} + h_C^2~\mbox{and}~ \delta_{n2} = \frac{1}{\sqrt{n \rho_n h_S}} +\frac{1}{\sqrt{n}} + h_S^2.
$$ 
\end{theorem}

\begin{theorem}
	\label{thm_beta_discrete}
	Suppose that Conditions~\ref{cond_eigen}--\ref{cond_bias_scalar} and \ref{cond_discrete_mixing}--\ref{cond_bandwidths} hold.  The following assertions hold as $n\rightarrow \infty:$\\
	(i) Let $\epsilon_n \rightarrow 0$ and $\epsilon_n^2n \rightarrow \infty$ as $n \rightarrow \infty.$ When $d$ is fixed, then 
    $$
    \|\widetilde{\beta}- \beta_0\|=O_P\big(\delta_{n1} + \delta_{n2}\big).
    $$
	(ii) When $d = \infty,$ if we further assume that $M\asymp \delta_{n1}^{-2/(2\alpha + 2\tau)} + \delta_{n2}^{-2/(2\alpha + 2\tau)},$ then
	$$
	\|\widetilde{\beta}- \beta_0\|^2 = O_P\Big\{M^{2\alpha +1}\big(\delta_{n1}^2 + \delta_{n2}^2\big) + M^{-2\tau +1}\Big\} = O_P\Big\{ \delta_{n1}^{\frac{2(2 \tau - 1)}{2\alpha+2\tau}} + \delta_{n2}^{\frac{2(2 \tau - 1)}{2\alpha+2\tau}}\Big\}.
	$$
\end{theorem}

{\bf Remarks.}  
(a) In the sparse case where $\rho_n$ is bounded, the $L_2$ rates of convergence for $\widetilde C_k$ and $\widetilde S_k$ in Theorem~\ref{thm_covariance_discrete} become $O_P(n^{-1/2}h_C^{-1}+h_C^2)$ and $O_P(n^{-1/2}h_S^{-1/2}+h_S^2),$ respectively, which are consistent to those yielded convergence
rates of one-dimensional and surface local linear smoothers for independent and sparsely sampled functional data \cite[]{zhang2016}.
When $\rho_n$ grows with $n,$ the convergence result reveals  interesting phase transition phenomena depending on the relative order of $\rho_n$ to $n.$ We use different rates of $\widetilde C_k$ ($k \geq 1$) to illustrate such phenomenon:
\begin{enumerate}
    \item[i.] When $\rho_n/n^{1/4} \rightarrow 0$ with $n^{1/4}h \rightarrow \infty,$ 
    $\|\widetilde C_k -C_k\|=O_P(n^{-1/2}\rho_n^{-1}h_C^{-1}+h_C^2);$
    \item[ii.] When $\rho_n \asymp n^{1/4}$ with $h_C \asymp n^{-1/4}$ or $\rho_n/n^{1/4}\rightarrow \infty$ with $h_C = o(n^{-1/4})$ and $h_C\rho_n \rightarrow \infty,$ $\|\widetilde C_k -C_k\|=O_P(n^{-1/2}).$
\end{enumerate}
As $\rho_n$ grows very fast, case~(ii) results in the root-$n$ rate, presenting that the theory for very dense curve time series falls in the parametric paradigm. As $\rho_n$ grows moderately fast, case~(i) corresponds to the rate faster than that for sparse data but slower than root-$n.$ The rates under cases~(i) and (ii) are respectively consistent to those of the estimated covariance function under categories of ``dense" and and ``ultra-dense" functional data \cite[]{zhang2016}. For $\widetilde S_k$ ($k \geq 1$), similar phase transition phenomenon occurs based on the ratio of $\rho_n$ to $n^{1/4}.$ 

(b) The $L_2$ rates of $\widetilde \beta_0$ in Theorem~\ref{thm_beta_discrete} are governed by dimensionality parameters ($n, \rho_n),$ bandwidth parameters $(h_C, h_S)$ and those internal parameters in part~(ii) of Theorem~\ref{thm_est_scalar} when $d=\infty$. There also exists the phase transition based on the relative order of $\rho_n$ to $n.$ For example, when $\rho_n$ is bounded and $d$ is fixed, the rate of $\widetilde \beta_0$ is  $O_P(n^{-1/2}h_C^{-1}+n^{-1/2}h_S^{-1/2}+h_C^2+h_S^2).$
When $\rho_n$ grows very fast with $\rho_n^{-1}=O(n^{-1/4})$ and suitable choices of $h_C, h_S,$ the rates of $\widetilde \beta_0$ are identical to those for fully observed functional predictors in Theorem~\ref{thm_est_scalar}.

\section{Empirical studies}
\label{sec.emp}

\subsection{Simulation study}
\label{subsec.sim}
In this section, we evaluate the finite sample performance of AGMM by a number of simulation studies. The observed predictor curves, $W_t(u), u \in [0,1],$ are generated from equation (\ref{curve.decomp}) with 
$$
X_t(u)=\sum_{j=1}^d\xi_{tj}\phi_j(u) ~~ \mbox{and}~~ e_t(u)=\sum_{j=1}^{10} \nu_{tj}\zeta_j(u),
$$
where 
$\{\xi_{tj}\}_{t=1}^n$ follows a linear AR(1) process with the coefficient $(-1)^j(0.9-0.5j/d)$. The slope functions are generated by $\beta_0(u)=\sum_{j=1}^d b_j \phi_j(u),$ where $b_j$'s take values from the first $d$ components in $(2, 1.6, -1.2, 0.8, -1, -0.6).$ We generate responses $Y_1, \dots, Y_n$ from equation (\ref{flm.scalar}), where $\varepsilon_t$ are independent $N(0,1)$ variables. Finally, we consider two different scenarios to generate $\{\phi_j(\cdot)\}_{j=1}^d,$ $\{\zeta_j(\cdot)\}_{j=1}^{10}$ and $\{\nu_{tj}\}_{n \times 10}.$

\textbf{Example~1}: 
This example is taken from \cite{bathia2010} with  $$\phi_j(u)=\sqrt{2}\cos(\pi j u), \quad \zeta_j(u)=\sqrt{2}\sin(\pi j u),$$ and the innovations $\nu_{tj}$ being independent standard normal variables.

We compare two versions of AGMM with three competing methods: covariance-based LS (CLS), covariance-based GMM (CGMM), autocovariance-based LS (ALS). The
three competing approaches are implemented as follows. In the first two methods, we perform eigenanalysis on the estimated covariance function $\widehat C_W,$ which converts the functional linear regression to the multiple linear regression, and then implement either LS or GMM. The truncated dimension was chosen such that the selected principal components can explain more than $90\%$ of the variation in the trajectory. We also tried the bootstrap method in \cite{hall2006} or to set a larger threshold level, e.g. $95\%.$ However neither approach performed well, so we do not report the results here. The third ALS method relies on the eigenanalysis on the estimated autocovariance-based $\widehat K$ and the subsequent implementation of LS. In a similar fashion to the difference between Base AGMM and AGMM, we refer to each of the unregularized method as the ``base" version. 

\begin{figure*}[t]
	\centering
	\includegraphics[width=5.00cm,height=5.00cm]{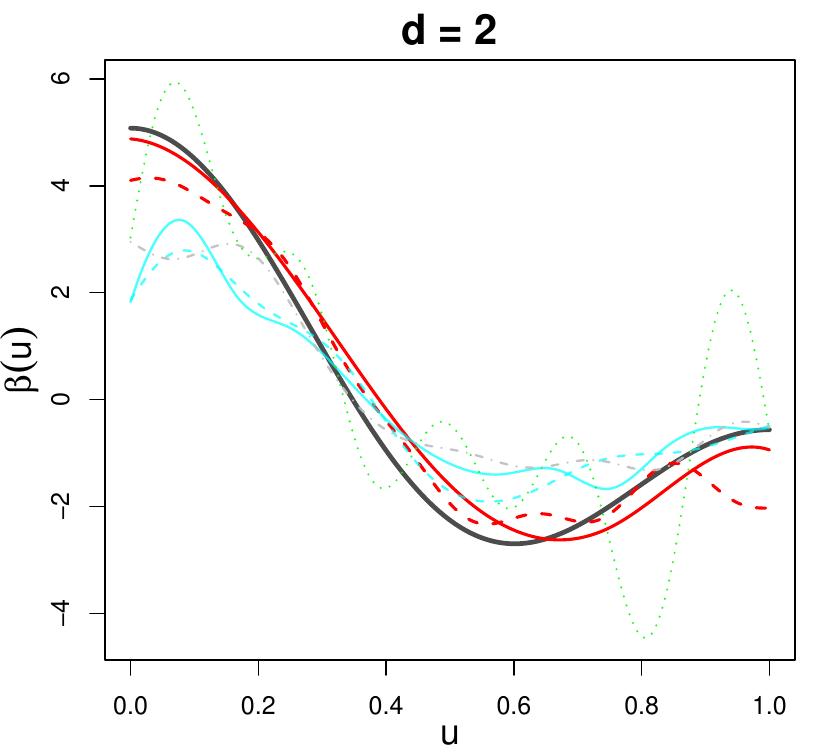}\includegraphics[width=5.00cm,height=5.00cm]{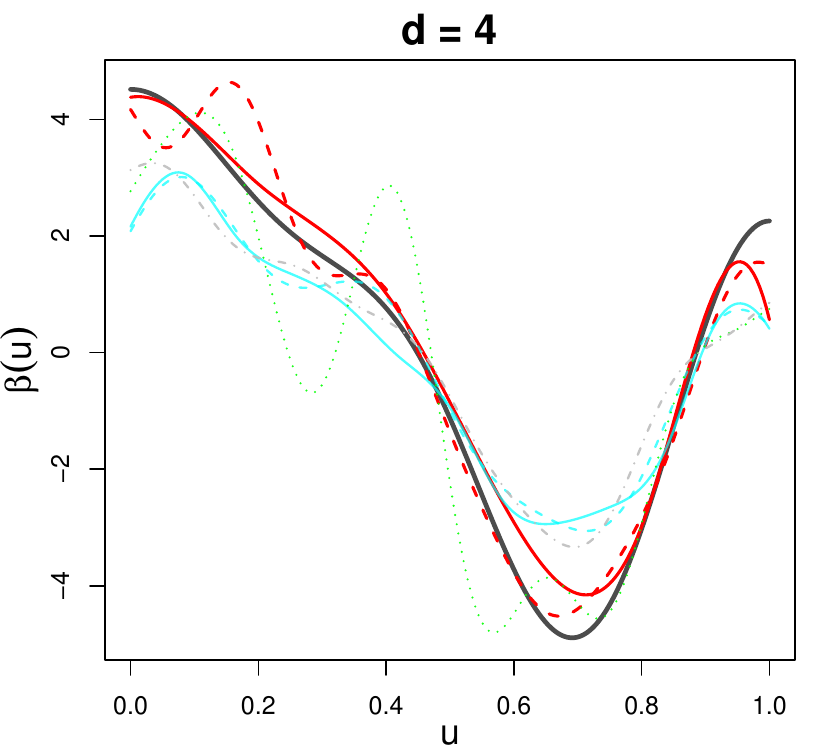}\includegraphics[width=5.00cm,height=5.00cm]{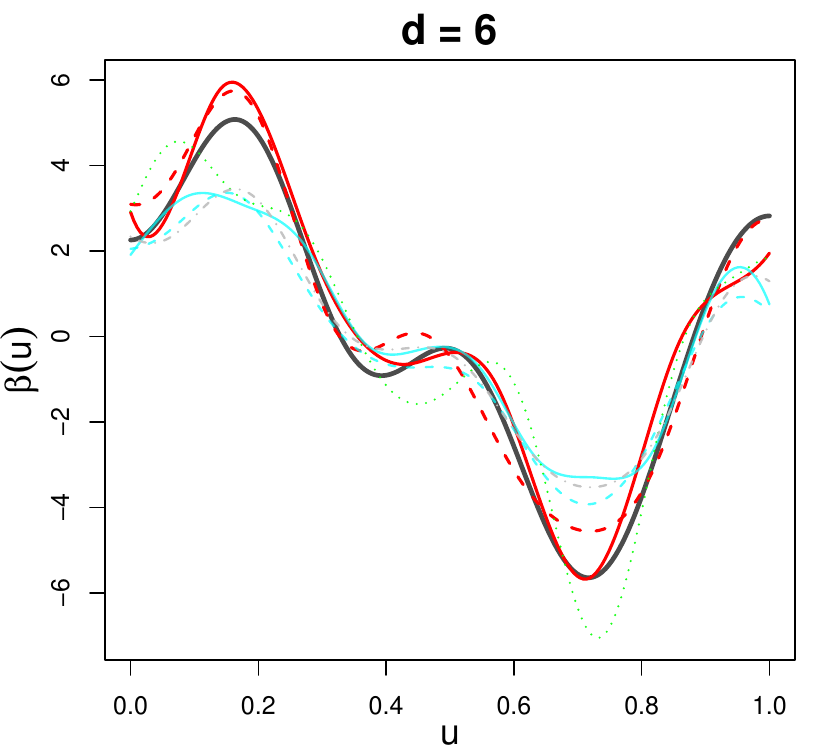}
	\caption{\label{sim.plot1}{Example~1 with $n=800$ and $d=2, 4, 6$: Comparison of true $\beta(\cdot)$ functions (black solid) with median estimates over 100 simulation runs for AGMM (red solid), Base AGMM (red dashed), CLS (cyan solid), Base CLS (cyan dashed), Base CGMM (green dotted) and Base ALS (gray dash-dotted).}}
\end{figure*}

The performance of four types of approaches are examined based on the mean integrated squared error for $\widehat \beta(u),$ i.e. $E[\int \{\widehat\beta(u)-\beta_0(u)\}^2du].$
We consider different settings with $d=2,4,6$ and $n=200, 400, 800,$ and ran each simulation 100 times. The regularized versions of CGMM and ALS did not give improvements in our simulation studies, so we do not report their results here. Figure~\ref{sim.plot1} provides a graphical illustration of the results for $n=800$ and $d=2,4,6.$ The black solid lines correspond to the true $\beta(u)$ from which the data were generated. The median most accurate estimate is also plotted for each of the competing methods. It is easy to see that the AGMM methods apparently provide the highest level of accuracy. The top part of Table~\ref{table.m1} reports numerical summaries for all simulation scenarios. We can observe that the advantage of AGMM over Base AGMM is prominent especially when either $d$ or $n$ is relatively small, while AGMM methods are superior to the competing methods when $n=400$ or $800.$ However, under the setting with $n=200$ and $d=4$ or $6$, the bootstrap test in Section~\ref{sec.tune} could not select $\widehat d$ very accurately, thus resulting in AGMM estimates inferior to some competitors. 

\begin{table}[ht]
	\caption{\label{table.m1} {\it Example~1}: The mean and standard error (in parentheses) of the mean integrated squared error for $\widehat \beta(u)$ over 100 simulation runs. The lowest values are in bold font.}
	\begin{center}
		\resizebox{6.2in}{!}{
			\begin{tabular}{ccc|cccccc}
				$\widehat d$& $n$ & $d$ & Base CLS & CLS & Base CGMM & Base ALS & Base AGMM & AGMM\tabularnewline
				\hline
				\multirow{9}{*}{Est}   &    \multirow{3}{*}{200}  & 2 & 1.320(0.026) & 1.315(0.025) & 2.215(0.099) & 1.619(0.044) & 1.187(0.052) & \bf{0.720(0.033)}\tabularnewline
				&  & 4 & 1.360(0.028) & \bf{1.340(0.028)} & 2.128(0.093) & 2.451(0.102) & 2.053(0.117) & 1.704(0.107)\tabularnewline
				&  & 6 & 1.337(0.030) & \bf{1.320(0.029)} & 1.912(0.102) & 2.150(0.092) & 1.847(0.098) & 1.612(0.072)\tabularnewline
				& \multirow{3}{*}{400}  & 2 & 1.184(0.018) & 1.181(0.019) & 1.891(0.090) & 1.338(0.026) & 0.772(0.034) & \bf{0.498(0.028)}\tabularnewline
				&   & 4 & 1.198(0.021) & 1.199(0.021) & 1.939(0.090) & 1.316(0.028) & 0.701(0.034) & \bf{0.584(0.034)}\tabularnewline
				&   & 6 & 1.159(0.023) & 1.154(0.022) & 1.519(0.087) & 1.323(0.034) & 0.824(0.045) & \bf{0.745(0.037)}\tabularnewline
				& \multirow{3}{*}{800}  & 2 & 1.159(0.012) & 1.158(0.012) & 1.792(0.080) & 1.161(0.013) & 0.346(0.013) & \bf{0.211(0.012)}\tabularnewline
				&   & 4 & 1.161(0.014) & 1.160(0.014) & 1.762(0.105) & 1.122(0.014) & 0.336(0.015) & \bf{0.247(0.012)}\tabularnewline
				&   & 6 & 1.123(0.014) & 1.122(0.014) & 1.297(0.091) & 1.119(0.016) & {\bf 0.348(0.016)} & 0.350(0.018)\tabularnewline
				\hline
				\multirow{9}{*}{True}   &        \multirow{3}{*}{200}  & 2 & 1.402(0.032) & 1.238(0.030) & 0.774(0.044) & 1.637(0.044) & 1.196(0.052) & \bf{0.718(0.033)}\tabularnewline
				&   & 4 & 1.365(0.030) & 1.191(0.029) & 0.924(0.056) & 1.515(0.043) & 1.214(0.071) & \bf{0.797(0.046)}\tabularnewline
				&  & 6 & 1.345(0.028) & 1.272(0.027) & \bf{1.150(0.065)} & 1.465(0.036) & 1.378(0.070) & 1.196(0.057)\tabularnewline
				&  \multirow{3}{*}{400}  & 2 & 1.226(0.019) & 1.145(0.019) & 0.503(0.027) & 1.336(0.026) & 0.772(0.034) & \bf{0.498(0.028)}\tabularnewline
				&   & 4 & 1.199(0.021) & 1.139(0.021) & 0.529(0.024) & 1.237(0.022) & 0.653(0.032) & \bf{0.488(0.029)}\tabularnewline
				&   & 6 & 1.166(0.023) & 1.139(0.022) & \bf{0.656(0.038)} & 1.170(0.023) & 0.726(0.039) & 0.704(0.042)\tabularnewline
				& \multirow{3}{*}{800}  & 2 & 1.174(0.012) & 1.136(0.012) & 0.269(0.011) & 1.161(0.013) & 0.346(0.013) & \bf{0.211(0.012)}\tabularnewline
				&   & 4 & 1.165(0.014) & 1.131(0.014) & 0.324(0.014) & 1.130(0.014) & 0.333(0.015) & 0.\bf{245(0.012)}\tabularnewline
				&   & 6 & 1.121(0.014) & 1.119(0.014) & \bf{0.323(0.016)} & 1.106(0.015) & 0.336(0.015) & 0.334(0.016)\tabularnewline
				\hline
				\vspace{-0.8cm}
			\end{tabular}
		}	
	\end{center}
\end{table}

To investigate the performance of AGMM after excluding the negative impact from the low accuracy of $\widehat d$ especially when $n=200,$ we also implement an ``oracle" version, which uses the true $d$ in the estimation. The numerical results are reported in the bottom part of Table~\ref{table.m1}. We can observe that GMM methods are superior to their LS versions, while CGMM slightly outperforms AGMM. These observations are due to the facts that, (i) top $d$ eigenvalues for $C_W$ and $K$ correspond to the same signal components in Example~1, (ii) GMM methods are capable of removing the impact from the noise term, (iii) the estimate $\widehat C_W$ in CGMM does not consider the functional error, while $\widehat K$ in AGMM would suffer from error accumulations. To better demonstrate the superiority of AGMM, we explore Example~2 below, where the covariance-based approach would fail to identify the signal components but its autocovariance-based version could. 

\textbf{Example~2}: We generate $\{\zeta_j(\cdot)\}_{j=1}^{10}$ from a $10$-dimensional orthonormal Fourier basis function, $\{\sqrt{2}\cos(2\pi j u), \sqrt{2}\sin(2\pi j u)\}_{j=1}^5,$ and set $\phi_j(u)=\zeta_j(u)$ for $j=1,\dots, d.$ The innovations $\nu_{tj}$ are independently sampled from $N(0,\sigma_j^2)$ with
$$\sigma_j^2=\begin{cases} (1/2)^{j-1}, \text{ for } j=1,\dots,6, \\
(2.6-0.1j)\times 1.1^{(d/2-3)}, \text{ for } j=7,\dots,10.\end{cases}$$

\begin{figure*}[ht]
	\centering
	\includegraphics[width=5.00cm,height=5.00cm]{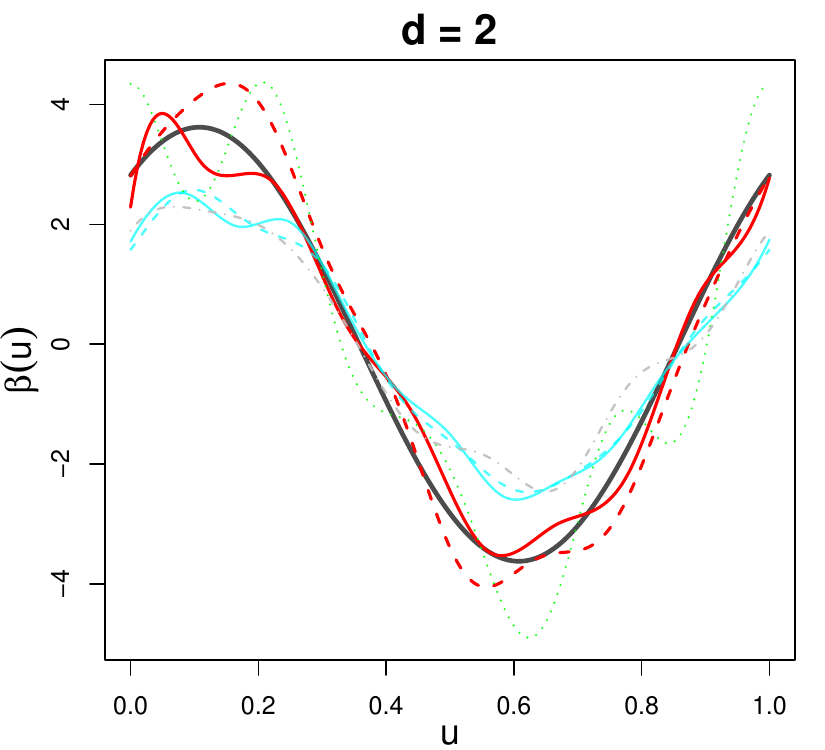}\includegraphics[width=5.00cm,height=5.00cm]{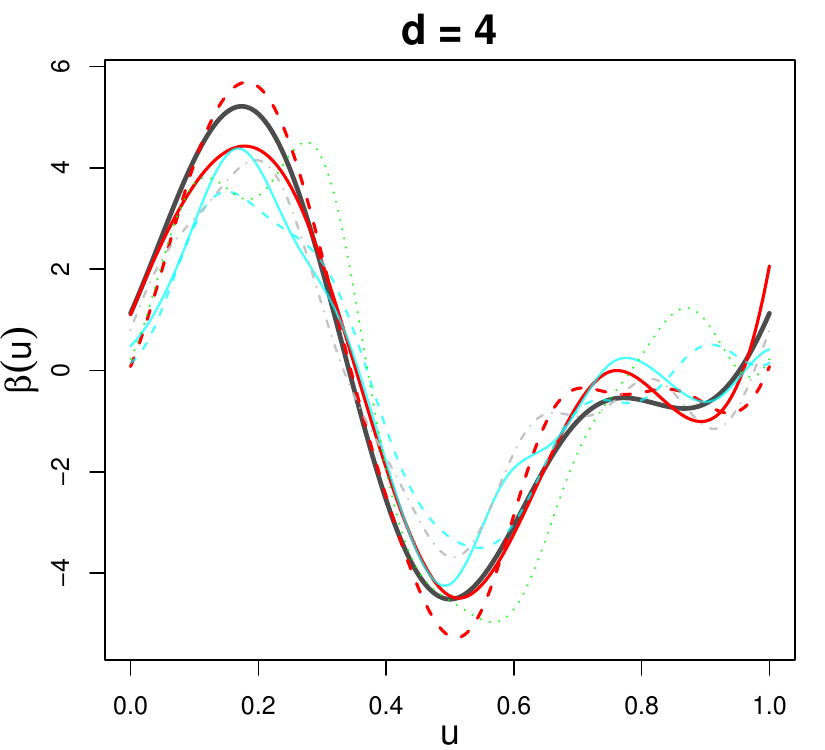}\includegraphics[width=5.00cm,height=5.00cm]{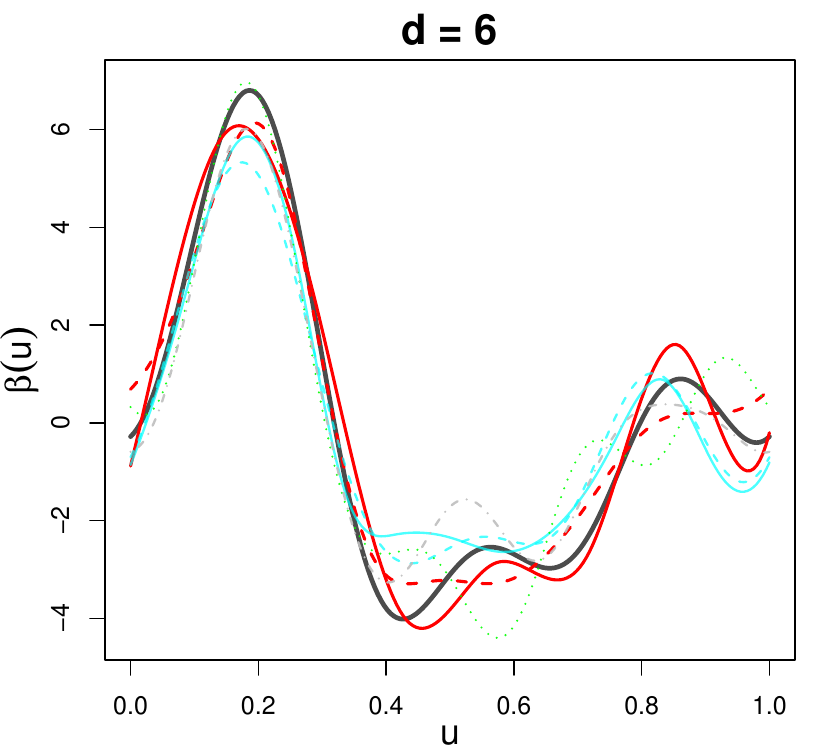}
	\caption{\label{sim.plot2}{Example~2 with $n=800$ and $d=2, 4, 6$: Comparison of true $\beta(\cdot)$ functions (black solid) with median estimates over 100 simulation runs for AGMM (red solid), Base AGMM (red dashed), CLS (cyan solid), Base CLS (cyan dashed), Base CGMM (green dotted) and Base ALS (gray dash-dotted).}}
\end{figure*}

In this example,  provided the fact that $\{\phi_j(\cdot)\}_{j=1}^{d}$ shares the common basis functions with the first $d$ elements in $\{\zeta_j(\cdot)\}_{j=1}^{10},$ we can calculate the variation in the trajectory explained by each of the 10 components under the population level. See Table~\ref{table.m2.eigen} 
of the Supplementary Material for details. Take $d=4$ as an illustrative example, the autocovariance-based methods can correctly identify the 4 signal components, while CLS and CGMM would mis-identify ``$7$" and ``$8$" as the signal components. Table~\ref{table.m2} gives numerical summaries under the ``oracle" scenario with true $d$ in the estimation. As we would expect, two versions of AGMM provide substantially improved estimates, while Base AGMM is outperformed by AGMM in most of the cases. Under the scenario that $\widehat d$ is selected by the bootstrap approach, Figure~\ref{sim.plot2} and Table~\ref{table.m2} provide the graphical and numerical results, respectively. We observe similar trends as in
Figure~\ref{sim.plot1} and Table~\ref{table.m1} with AGMM methods providing highly significant improvements over all the competitors.

\begin{table}[ht]
	\caption{\label{table.m2} {\it Example~2}: The mean and standard error (in parentheses) of the mean integrated squared error for $\widehat \beta(u)$ over 100 simulation runs.
		The lowest values are in bold font.}
	\begin{center}
		\resizebox{6.2in}{!}{
			\begin{tabular}{ccc|cccccc}
				$\widehat d$& $n$ & $d$ & Base CLS & CLS & Base CGMM & Base ALS & Base AGMM & AGMM\tabularnewline
				\hline
				\multirow{9}{*}{True}  & \multirow{3}{*}{400}  & 2 & 1.591(0.059) & 0.990(0.046) & 1.118(0.078) & 1.165(0.030) & 0.599(0.038) & \bf{0.262(0.026)}\tabularnewline
				&   & 4 & 2.026(0.066) & 1.590(0.070) & 2.310(0.112) & 0.972(0.033) & 0.686(0.041) & \bf{0.448(0.034)}\tabularnewline
				&   & 6 & 2.310(0.069) & 1.932(0.077) & 2.722(0.104) & 0.938(0.035) & 0.825(0.042) & \bf{0.676(0.048)}\tabularnewline
				& \multirow{3}{*}{800}  & 2 & 1.377(0.051) & 0.940(0.038) & 0.884(0.085) & 0.994(0.019) & 0.337(0.020) & \bf{0.138(0.010)}\tabularnewline
				&   & 4 & 1.934(0.051) & 1.526(0.054) & 2.268(0.105) & 0.685(0.016) & 0.318(0.016) & \bf{0.208(0.013)}\tabularnewline
				&   & 6 & 2.160(0.056) & 1.872(0.055) & 2.859(0.138) & 0.575(0.015) & \bf{0.339(0.017)} & 0.364(0.020)\tabularnewline
				& \multirow{3}{*}{1200}  & 2 & 1.294(0.053) & 0.980(0.048) & 0.750(0.081) & 0.900(0.013) & 0.203(0.011) & {\bf 0.080(0.005)}\tabularnewline
				&  & 4 & 1.959(0.053) & 1.524(0.058) & 2.426(0.121) & 0.582(0.009) & 0.167(0.008) & {\bf 0.124(0.006)}\tabularnewline
				&   & 6 & 2.270(0.048) & 2.002(0.050) & 3.092(0.113) & 0.494(0.011) & {\bf 0.217(0.010)} & 0.248(0.010)\tabularnewline
				\hline
				\multirow{9}{*}{Est} &\multirow{3}{*}{400}  & 2 & 0.817(0.012) & 0.818(0.012) & 0.980(0.059) & 1.141(0.026) & 0.575(0.030) & \bf{0.248(0.018)}\tabularnewline
				&  & 4 & 1.037(0.043) & 0.725(0.036) & 1.319(0.070) & 1.097(0.038) & 0.773(0.042) & \bf{0.584(0.038)}\tabularnewline
				&  & 6 & 0.913(0.041) & \bf{0.811(0.038)} & 1.305(0.068) & 1.164(0.050) & 0.999(0.051) & 0.955(0.053)\tabularnewline
				& \multirow{3}{*}{800}  & 2 & 0.795(0.010) & 0.795(0.010) & 0.899(0.055) & 0.989(0.019) & 0.333(0.020) & \bf{0.138(0.009)}\tabularnewline
				&   & 4 & 1.093(0.033) & 0.768(0.035) & 1.471(0.065) & 0.682(0.016) & 0.319(0.016) & \bf{0.212(0.013)}\tabularnewline
				&   & 6 & 0.859(0.041) & 0.809(0.039) & 1.139(0.061) & 0.571(0.016) & \bf{0.335(0.017)} & 0.369(0.020)\tabularnewline
				& \multirow{3}{*}{1200}  & 2 & 0.779(0.007) & 0.780(0.007) & 0.747(0.044) & 0.898(0.012) & 0.205(0.012) & \bf{0.079(0.005)}\tabularnewline
				&  & 4 & 1.055(0.026) & 0.815(0.032) & 1.344(0.052) & 0.580(0.009) & 0.166(0.008) & \bf{0.130(0.007)}\tabularnewline
				&   & 6 & 0.813(0.029) & 0.808(0.029) & 1.159(0.058) & 0.492(0.011) & \bf{0.216(0.011)} & 0.243(0.009)\tabularnewline
				\hline
				\vspace{-0.8cm}
			\end{tabular}
		}	
	\end{center}
\end{table}

\textbf{Example~3}:
We use this example to demonstrate the sample performance of our proposed kernel smoothing approach to handle partially observed functional predictors. In each simulated scenario, we first generate $\{W_t(\cdot)\}$ and $\{e_t(\cdot)\}$ in the same way as Example~2 and then generate the observed values $Z_{ti}$ from equation (\ref{curve.patial}), where time points $U_{ti}$ and errors $\eta_{ti}$ are randomly sampled from Uniform$[0,1]$ and $N(0, 0.5^2),$ respectively. We consider simulation settings $d=2,4,6,$ $n=400, 800, 1200$ and $m_t=10, 25, 50, 100,$ changing from sparse to moderately dense to very dense measurement schedules. In each case, the optimal bandwidth parameters, $h_C, h_S$, are selected by the $10$-fold cross-validation in \cite{rubin2020} and $\widehat d$ is chosen so that the first $\widehat d$ eigenvalues explains over 95\% of the total variation. Table~\ref{table.m3} reports numerical summaries for all 36 cases. Several conclusions can be drawn. First, for each $d,$ the estimation accuracy is improved as $n$ and $m_t$ increase. Second, as curves are very densely observed, e.g. $m_t=100$, our proposed smoothing approach enjoys similar performance with AGMM in Table~\ref{table.m2}, providing empirical evidence to support our remark for Theorem~\ref{thm_beta_discrete} about the same convergence rate between very densely observed and fully observed functional scenarios. 

\begin{table}[ht]
	\caption{\label{table.m3} {\it Example~3}: The mean and standard error (in parentheses) of the mean integrated squared error for $\widehat \beta(u)$ over 100 simulation runs.}
	\begin{center}
		\resizebox{4.2in}{!}{
			\begin{tabular}{cc|cccc}
				$n$ & $d$ & $m_t=10$ & $m_t=25$ & $m_t=50$ & $m_t=100$ \tabularnewline
				\hline
				\multirow{3}{*}{400}
				& 2 & 0.906(0.052) & 0.374(0.019) & 0.296(0.015) & 0.227(0.011)\tabularnewline
				& 4 & 1.238(0.046) & 0.637(0.027) & 0.593(0.045) & 0.395(0.020)\tabularnewline
				& 6 & 1.168(0.051) & 1.092(0.031) & 0.906(0.028) & 0.721(0.027)\tabularnewline
				\multirow{3}{*}{800}
				& 2 & 0.571(0.030) & 0.194(0.009) & 0.155(0.008) & 0.142(0.007)\tabularnewline
				& 4 & 0.804(0.030) & 0.375(0.015) & 0.329(0.023) & 0.231(0.010)\tabularnewline
				& 6 & 1.130(0.039) & 0.835(0.029) & 0.481(0.019) & 0.360(0.013)\tabularnewline
				\multirow{3}{*}{1200}
				& 2 & 0.317(0.017) & 0.145(0.007) & 0.124(0.006) & 0.107(0.005)\tabularnewline
				& 4 & 0.632(0.025) & 0.226(0.008) & 0.214(0.013) & 0.150(0.007)\tabularnewline
				& 6 & 1.043(0.031) & 0.505(0.016) & 0.311(0.010) & 0.269(0.009)\tabularnewline
				\hline
				\vspace{-0.8cm}
			\end{tabular}
		}	
	\end{center}
\end{table}

\subsection{Real data analysis}
\label{subsec.real}
In this section, we illustrate the proposed AGMM using a public financial dataset. The dataset was downloaded from \url{https://wrds-web.wharton.upenn.edu/wrds} and consists of one-minute resolution prices of Standard \& Poor's 500 index and inclusive stocks from $n=251$ trading days in year~2017. The trading time (9:30-16:00) is then converted to minutes, $u \in [0,390].$ Let $P_t(u_j)$ ($t=1, \dots, n,j=1, \dots, 390$) be the price of a financial asset at the $j$-th minute after the opening time on the $t$-th trading day. Denote the {\it cumulative intraday return} (CIDR) trajectory, in percentage, by  $r_{t}(u_j)=100\big[\log\{P_t(u_j)\}-\log\{P_t(u_1)\}\big]$ \cite[]{horvath2014}. Let $r_{m,t}(u)$ be the CIDR curves of the Standard \& Poor's 500 index. 

We extend the standard {\it capital asset pricing model} (CAPM) [Chapter~5 of \cite{Bcampbell1997}] to the functional domain by considering the functional linear regression with functional errors-in-predictors as follows
\begin{equation}
\label{fcapm}
y_t = \alpha + \int x_{t}(u)\beta(u)du +  \varepsilon_t, ~~ r_{m,t}(u) = x_{t}(u) + e_{t}(u), ~t=1, \dots,n, ~u \in [0,390],
\end{equation}
where $x_t(\cdot)$ and $e_t(\cdot)$ represent the signal and error components in $r_{m,t}(\cdot),$ respectively, and 
$y_t$ is the intraday return of a specific stock on the $t$-th trading day. 
Note that the slope parameter in the classical CAPM explains how strongly an asset return depends on the market portfolio. Analogously, $\beta(\cdot)$ in functional CAPM in (\ref{fcapm}) can be understood as the functional sensitivity measure of an asset return to the market CIDR trajectory. 

\begin{figure*}[t]
	\centering
	\includegraphics[width=5.00cm,height=5.00cm]{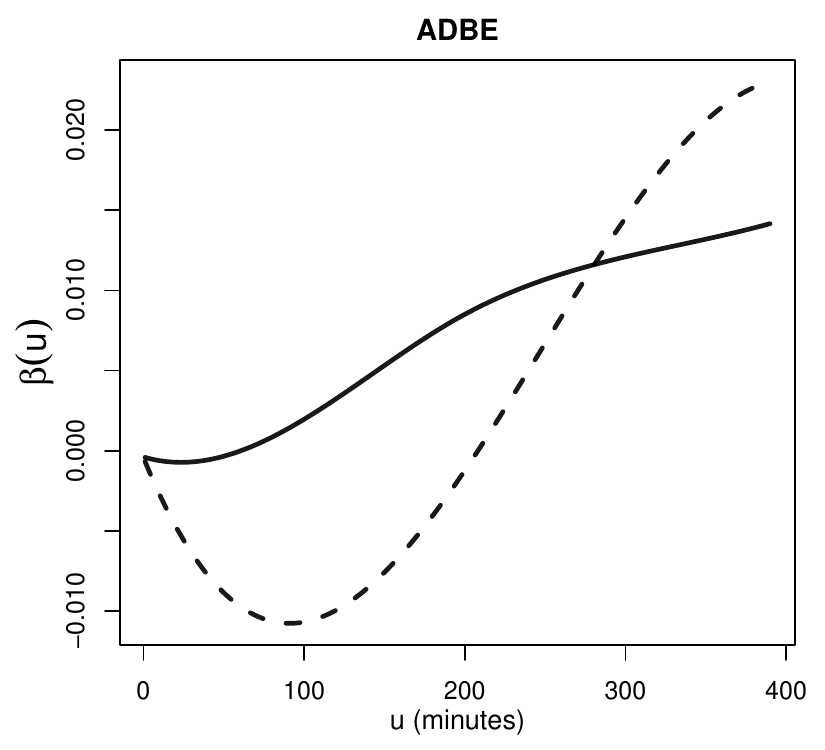}\includegraphics[width=5.00cm,height=5.00cm]{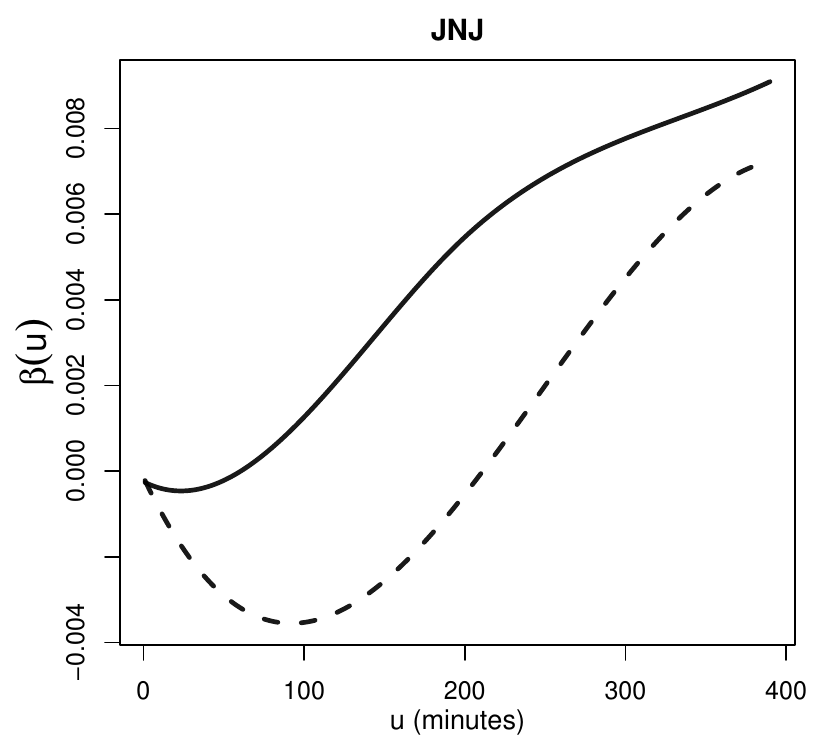}\includegraphics[width=5.00cm,height=5.00cm]{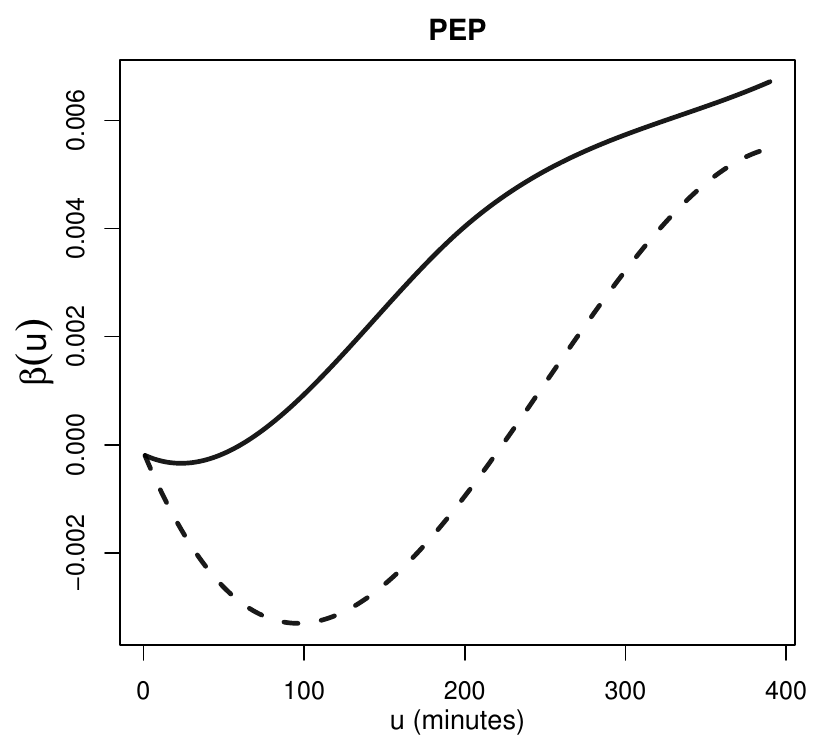}
	\caption{\label{real.plot}{Estimated $\beta(\cdot)$ curves for AGMM (solid) and CLS (dashed)
		}.}
\end{figure*}

Figure~\ref{real.plot} plots the estimated $\beta(\cdot)$ functions using both AGMM and CLS for three large-cap-sector stocks, Adobe (ADBE), Johnson \& Johnson (JNJ) and PepsiCo (PEP). 
A few trends are apparent. First, the AGMM estimates place more positive weights as $u$ increases. This result seems reasonable given the fact that the daily most recent market price would contain the most information about the stock's closing price. Second, the CLS estimates first dip in the mid-morning and then start to increase until the end of the trading day. In general, the shapes of the estimated $\beta(\cdot)$ functions by either AGMM or CLS are quite similar across the three stocks.

To formulate a prediction problem, we treat CIDR trajectories of the same stock as that in (\ref{fcapm}) up to current time $T<390$ as $r_{y,t}(u), u \in [0,T],$ where, e.g., $T=375$ corresponds to 15 minutes prior to the closing time of the trading day. Then we construct the same functional linear model as (\ref{fcapm}) by replacing $r_{m,t}(\cdot)$ with $r_{y,t}(\cdot).$ To judge which method produces superior predictions, we implement a rolling procedure to calculate the mean squared prediction error (MSPE) for $H=30$ days. Specifically, for each $h=H, H-1, \dots, 1,$ we treat $\{y_{n-h+1}, r_{y,n-h+1}\}$ as a testing set, implementing each fitting method on the training set of $\{(y_{t}, r_{y,t}): t=1, \dots, n-h\},$ calculate the squared error between $y_{n-h+1}$ and its predicted value, and repeat this procedure $H$-times to compute the MSPE. We calculate the MSPEs over a grid of $(d,J)$ values and choose the pair with the lowest error. We also include the prediction errors from the null model, using the mean of the training response to predict the test response. The resulting MSPEs, for various values of $T$ and the same three stocks, are provided in Table~\ref{table.real}. It is easy to observe that the prediction accuracy for AGMM and CLS improves as $T$ approaches to $390$ and AGMM significantly outperforms two competitors in almost all settings.

\begin{table}[ht]
	\caption{\label{table.real} Mean squared prediction errors up to different current times, $T=$ 330, 345, 360, 375, 380 and 385 minutes, for AGMM and two competing methods. All entries have been multiplied by 10 for formatting reasons. The lowest MSPE for each value of T is bolded.}
	\begin{center}
		\resizebox{4.8in}{!}{
			\begin{tabular}{cc|cccccc}
				Stock& Method & $u \leq 330$ & $u \leq 345$ & $u \leq 360$ & $u \leq 375$ & $u \leq 380$ & $u \leq 385$  \tabularnewline
				\hline
				\multirow{3}{*}{ADBE} & AGMM  & 1.276 &  {\bf 1.179} & {\bf 0.983} & {\bf 0.852} & {\bf 0.800} & {\bf 0.728} \tabularnewline 
				&  CLS &  {\bf 1.272} & 1.186 & 1.094 & 0.991 & 0.949 & 0.895 \tabularnewline
				&  Mean & 12.224 & 12.224& 12.224 & 12.224 & 12.224 &  12.224\tabularnewline
				\hline
				\multirow{3}{*}{JNJ} & AGMM  & {\bf 0.419} & {\bf 0.305} & {\bf 0.279} & {\bf 0.254} & {\bf 0.243} & {\bf 0.226} \tabularnewline
				&  CLS & 0.583 & 0.496 & 0.419 & 0.352 & 0.330 & 0.306  \tabularnewline
				&  Mean & 3.077 & 3.077&3.077 &3.077 & 3.077&  3.077\tabularnewline
				\hline
				\multirow{3}{*}{PEP} & AGMM  &  {\bf 0.749}& {\bf 0.659} & {\bf 0.557} & {\bf 0.466} & {\bf 0.429} & {\bf 0.384} \tabularnewline 
				&  CLS & 0.781 & 0.687 & 0.596& 0.502 & 0.468 & 0.429 \tabularnewline
				&  Mean & 2.956 & 2.956& 2.956& 2.956& 2.956&  2.956\tabularnewline
				\hline
				\vspace{-0.8cm}
			\end{tabular}
		}	
	\end{center}
\end{table}

\subsection*{Acknowledgements.} We are grateful to the editor, the associate editor and two referees for their insightful comments, which have led to significant improvement of our paper.

\appendix
\section{Appendix}
\label{SM}
Appendices~\ref{pf.thm1} and \ref{ap.pf.partial} contain proofs of Theorem~\ref{thm_est_scalar} and Theorems~\ref{thm_covariance_discrete}--\ref{thm_beta_discrete}. The proofs of Theorem~\ref{thm_est_func} and all technical lemmas are in the Supplementary Material.

\subsection{Proof of Theorem~\ref{thm_est_scalar}}
\label{pf.thm1}

\subsubsection{Proof of Theorem~\ref{thm_est_scalar}~(i)}



Define $\widecheck K(u,v)=\sum_{j=1}^d \widehat\theta_j \widehat\psi_j(u)\widehat\psi_j(v)$
and 
$
K^{-1}(u,v)=\sum_{j=1}^d \theta_j^{-1} \psi_j(u)\psi_j(v).
$
Let $\widecheck \beta(u)=\int_{\cU}\widecheck K^{-1}(u,v)\widehat R(v) dv.$ For a large $\delta>0,$ by Lemma~\ref{lemma.d}, 
we have
\begin{eqnarray*}
	P\big(n^{1/2}\|\widehat\beta-\beta_0\|>\delta\big) &=&
	P\big(n^{1/2}\|\widehat\beta-\beta_0\|>\delta,\widehat d=d\big) + P\big(n^{1/2}\|\widehat\beta-\beta_0\|>\delta,\widehat d \neq d\big)\\
	&\leq& P\big(n^{1/2}\|\widecheck\beta-\beta_0\|>\delta,\widehat d=d\big) +P\big(\widehat d \neq d\big)\\
	&\leq&P\big(n^{1/2}\|\widecheck\beta-\beta_0\|>\delta\big)+ o(1),
\end{eqnarray*}
which means that, to prove $ n^{1/2}\|\widehat\beta-\beta_0\| = O_P(1)$, it suffices to show that 
$
\|\widecheck\beta-\beta_0\|=O_P(n^{-1/2}).
$
It is easy to show that 
\begin{equation}
\label{beta.bd}
\|\widetilde\beta-\beta_0\| \leq \|\widecheck K^{-1}-K^{-1}\|_{\cS}\|\widehat R\| + \|K^{-1}\|_{\cS}\|\widehat R-R\|.    
\end{equation}
Then it follows from Lemmas~\ref{cond_moment_scalar}, \ref{lemma.R.bd} and \ref{lemma.Kinv} 
that $\|\widecheck\beta-\beta_0\|= O_P(n^{-1/2}).$ 


\subsubsection{Proof of Theorem~\ref{thm_est_scalar}~(ii)}

Without any ambiguity, write $\langle q,K \rangle$, $\langle K,q \rangle$ and $ \langle p, \langle K, q\rangle \rangle$ for 
$$
\int_\cU K(u,v) q(u)du, \int_\cU K(u,v) q(v)dv
~~ and ~~ \int_{\cU} \int_{\cU} K(u,v) p(u)q(v)du dv,
$$
respectively. In  Lemma~\ref{eigen.expansion}, 
we give expressions for $\widehat \theta_j - \theta_j$ and $\widehat \psi_j - \psi_j$ for $j \ge 1$.


Let $\beta_M(u)=\sum_{j=1}^{M}\theta_j^{-1} \langle \psi_j,R \rangle\psi_j(u).$ By the triangle inequality, we have 
\begin{eqnarray}
\label{triangle}
\|\widehat \beta -\beta_0\|^2 \le \|\widehat \beta-\beta_M\|^2 + \|\beta_M-\beta_0\|^2.
\end{eqnarray}
By (\ref{beta.sol.inf}) and orthonormality of $\{\psi_j(\cdot)\}$, we have 
$
\|\beta_M-\beta_0\|^2 = \sum_{j=M+1}^{\infty}\theta_j^{-2} \langle \psi_j,R \rangle^2.
$
It follows from Condition~\ref{cond_bias_scalar} and some specific calculations that
\begin{eqnarray}
\label{biasrate}
\|\beta_M-\beta_0\|^2 = 
\sum_{j=M+1}^{\infty} b_j^2 \leq C\sum_{j=M+1}^{\infty}j^{-2\tau} = O(M^{-2\tau + 1}).
\end{eqnarray}

Next we will show the convergence rate of $\|\widehat \beta - \beta_M\|^2$. Observe that 
\begin{eqnarray*}
	\widehat \beta(u) - \beta_M(u) 
	&=& \sum_{j=1}^{M} \big (\widehat \theta_j^{-1} - \theta_j^{-1} \big)\langle \psi_j,R \rangle \widehat \psi_j(u) + \sum_{j=1}^{M} \widehat \theta_j^{-1} \big(\langle \widehat \psi_j, \widehat R \rangle - \langle \psi_j, R \rangle\big) \widehat \psi_j(u) \\
	&& \hskip 1cm + \sum_{j=1}^{M} \theta_j^{-1} \langle  \psi_j, R \rangle \big\{\widehat\psi_j(u) - \psi_j(u)\big\}.
\end{eqnarray*}
Then we have
\begin{eqnarray}
\|\widehat \beta - \beta_M\|^2
&\le& 3\sum_{j=1}^{M} \big (\widehat \theta_j^{-1} - \theta_j^{-1} \big)^2 \langle \psi_j,R \rangle^2  + 3\sum_{j=1}^{M} \widehat \theta_j^{-2} \big(\langle \widehat \psi_j, \widehat R \rangle - \langle \psi_j, R \rangle\big)^2 \nonumber \\
&& \hskip 1cm + 3 M\sum_{j=1}^{M} \theta_j^{-2}\langle  \psi_j, R \rangle^2 \big\|\widehat\psi_j - \psi_j\big\|^2 \nonumber \\
& = & 3I_{n1} + 3I_{n2} + 3I_{n3}.
\end{eqnarray}

Let $\widehat \Delta = \|\widehat K - K\|_\cS$ and $\Omega_M = \{ 2 \widehat \Delta \le \delta_M \} $. On the event $\Omega_M$, we can see that $\sup_{j \le M}|\widehat \theta_j - \theta_j| \le \theta_M/2$, which implies that $2^{-1}\theta_j \le \widehat \theta_j \le 2 \theta_j$. Moreover, we can show that 
$P(\Omega_M) \to 1$ since $n^{1/2} \delta_M \to \infty$ as $n \to \infty$. Hence it suffices to work with bounds that are established under the event $\Omega_M$. 

Provided that event $\Omega_M$ holds, it follows from $\sup_{j \ge 1 }|\widehat \theta_j - \theta_j| = O_P(n^{-1/2})$ in Lemma~\ref{lemma.asy.eigen}(i) 
and some calculations that
\begin{eqnarray*}
	I_{n1} \le 4\sum_{j=1}^{M} \big (\widehat \theta_j - \theta_j \big)^2 \theta_j^{-4}\langle \psi_j,R \rangle^2 = 4 \sum_{j=1}^{M} \theta_j^{-2}b_j^2\big (\widehat \theta_j - \theta_j \big)^2 = O_P\Big(n^{-1}\sum_{j=1}^{M} \theta_j^{-2}b_j^2 \Big). 
\end{eqnarray*} 
By Conditions~\ref{cond_eigen}--\ref{cond_bias_scalar}, we have
\begin{eqnarray}
I_{n1} = O_P(n^{-1}) \cdot \Big(\sum_{j=1}^M j^{2 \alpha - 2 \tau}\Big) 
= O_P(n^{-1})\cdot \big(M + M^{2 \alpha - 2 \tau + 1}\big)
= o_P\big(n^{-1}M^{2 \alpha +1}\big).
\end{eqnarray}

Consider the term $I_{n3}.$ By $\|\widehat\psi_j - \psi_j\| = O_P\big(j^{1+\alpha} n^{-1/2}\big)$ in Lemma~\ref{lemma.asy.eigen}(iii) 
and Condition~\ref{cond_bias_scalar}, we obtain that
\begin{eqnarray}
I_{n3} \le M \sum_{j=1}^{M} b_j^2 \big\|\widehat\psi_j - \psi_j\big\|^2 = O_P\big( n^{-1}M^{2 - 2\tau + 2\alpha + 2}\big) = O_P\Big(n^{-1}M^{2\alpha + 1}\big),
\end{eqnarray}
where the last equality comes from $\alpha > 1$ and $2\alpha - 2 \tau + 4 \le 2 \alpha + 1$ implied by Condition~\ref{cond_bias_scalar}.

Consider the term $I_{n2}$. On the event $\Omega_M$, we have that
\begin{eqnarray}
I_{n2} &\le&  4\sum_{j=1}^{M}  \theta_j^{-2} \big(\langle \widehat \psi_j, \widehat R \rangle - \langle \psi_j, R \rangle\big)^2\nonumber\\
& \le & 12 \sum_{j=1}^{M}  \theta_j^{-2} \Big(\langle \widehat \psi_j - \psi_j, R \rangle^2 + \langle \psi_j, \widehat R - R \rangle^2 + \langle \widehat \psi_j - \psi_j, \widehat R - R \rangle^2 \Big)\nonumber\\
&\le &12 \sum_{j=1}^{M}  \theta_j^{-2} \Big ( \langle \widehat \psi_j - \psi_j, R \rangle^2 + \|\widehat R - R\|^2 + \|\widehat \psi_j - \psi_j\|^2\|\widehat R - R \|^2 \Big),
\label{bd.I2}
\end{eqnarray}
where the last inequality comes from orthonormality of $\{\psi_j(\cdot)\}$ and Cauchy-Schwarz inequality.
By Lemma~\ref{eigen.expansion} 
and some calculations, we can represent the term  $\langle \widehat \psi_j - \psi_j, R \rangle$ as
\begin{eqnarray*}
	\langle \widehat \psi_j - \psi_j, R \rangle = R_{j1} + R_{j2},
\end{eqnarray*}
where $R_{j1} =\sum_{k: k\neq j} \theta_k b_k (\widehat \theta_j - \theta_k)^{-1} \langle\widehat \psi_j, \langle \widehat K - K, \psi_k\rangle \rangle$ and $R_{j2} = \theta_j b_j \langle \widehat \psi_j - \psi_j, \psi_j\rangle$.
It follows from Condition~\ref{cond_eigen}--\ref{cond_bias_scalar}, Lemma~\ref{lemma.asy.eigen} 
and Cauchy-Schwarz inequality that
\begin{equation}
\label{I20.bd}
\sum_{j=1}^M \theta_j^{-2} R_{j2}^2 = O_P(n^{-1})\cdot \Big(\sum_{j=1}^M j^{-2 \tau + 2 \alpha + 2}\Big)= o_P\big(n^{-1}M^{2 \alpha +1}\big).
\end{equation}
Note that on the event $\Omega_M$, $|\widehat \theta_j - \theta_j| \le 2^{-1} |\theta_j - \theta_k|$ for  $j= 1,\ldots,k-1,k+1,\dots, M$ and hence $|\widehat \theta_j - \theta_k| \ge 2^{-1}|\theta_j - \theta_k|$. If we can show that 
\begin{eqnarray}
\label{bound_R1}
\sup_{j \ge 1}(\theta_j^2 j^{2\alpha})^{-1}\sum_{k: k\neq j} \theta_k^2 b_k^2 (\theta_j - \theta_k)^{-2} = O(1),
\end{eqnarray}
then, by Condition~\ref{cond_bias_scalar}, Lemma~\ref{lemma.asy.eigen} 
and on the event $\Omega_M$, we have
\begin{eqnarray}
\sum_{j=1}^M \theta_j^{-2} R_{j1}^2 
&\le& 4\sum_{j=1}^M \theta_j^{-2} \sum_{k: k\neq j} \theta_k^2 b_k^2 (\theta_j - \theta_k)^{-2} \|\widehat K - K\|_\cS^2 \nonumber\\
&=& O_P(n^{-1}) \cdot \sum_{j=1}^M \theta_j^{-2}\theta_{j}^2 j^{2\alpha} = O_P(n^{-1}M^{2\alpha + 1 }).
\label{I21.bd}
\end{eqnarray}

We turn to prove (\ref{bound_R1}) as follows. Denote $[j/2]$ by the largest integer less than $j/2$. Then
$$
\sum_{k: k\neq j} \theta_k^2 b_k^2 (\theta_j - \theta_k)^{-2} = 
\left(\sum_{k = 2 (j+1)}^{\infty} + \sum_{k = [j/2]+1, k \neq j}^{k = 2 j+1}+ \sum_{k = 1}^{ [j/2]}\right)\theta_k^2 b_k^2 (\theta_j - \theta_k)^{-2}.
$$
Observe that for $k \ge 2(j+1)$, 
$$
\theta_j - \theta_k = \sum_{s = j}^{k-1}(\theta_{s} - \theta_{s+1}) 
\ge c \int_{j+1}^{2(j+1)} s^{-\alpha - 1}ds = -\frac{c}{\alpha} s^{-\alpha}\Big|_{j+1}^{2(j+1)} \ge \frac{c}{2\alpha} 2^{-\alpha} j^{-\alpha},
$$
and for $[j/2]+2 \le k \le 2j + 1$ but $k \neq j$,
$$
|\theta_j - \theta_k| \ge \max(\theta_{j} - \theta_{j+1},\theta_{j-1} - \theta_{j}) \ge c j^{-\alpha -1}.
$$
Therefore,
\begin{eqnarray*}
	(\theta_j^2 j^{2\alpha})^{-1}\sum_{k=2(j+1)}^\infty \theta_k^2 b_k^2 (\theta_j - \theta_k)^{-2} &=& O(1)\cdot j^{2\alpha-2\tau} \sum_{k=2(j+1)}^\infty \theta_k^2=  O(1),\\
	(\theta_j^2 j^{2\alpha})^{-1} \sum_{k=[j/2]+1}^{2j+1} \theta_k^2 b_k^2 (\theta_j - \theta_k)^{-2} &\leq& 
	(\theta_j^2 j^{2\alpha})^{-1} \sum_{k=[j/2]+1}^{2j+1} 2\big\{\theta_j^2+(\theta_j-\theta_k)^2\big\}b_k^2(\theta_j-\theta_k)^{-2}\\
	&=& O(1)\cdot \theta_j^{-2}j^{-2\alpha}(1 + \theta_j^2j^{2\alpha+3-2\tau})=  O(1),\\
	(\theta_j^2 j^{2\alpha})^{-1} \sum_{k=1}^{[j/2]} \theta_k^2 b_k^2 (\theta_j - \theta_k)^{-2} &\leq& O(1)\sum_{k=1}^{[j/2]} \theta_k^2 b_k^2 (\theta_{k} - \theta_{2k})^{-2} = O(1)\cdot \theta_1^2 j^{2\alpha - 2\tau+1} =  O(1),
\end{eqnarray*}
uniformly in $j$. Then (\ref{bound_R1}) follows. 

Moreover, it follows from Condition~\ref{cond_eigen}, Lemmas~\ref{lemma.asy.eigen}--\ref{lemma.R.bd} 
that
\begin{equation}
\label{I22.bd}
\sum_{j=1}^{M}  \theta_j^{-2} \|\widehat R - R\|^2 = O_P(n^{-1} M^{2 \alpha + 1} ) ~ \mbox{and} ~
\sum_{j=1}^{M}  \theta_j^{-2} \|\widehat \psi_j - \psi_j \|^2\|\widehat R - R\|^2 = O_P(n^{-2}M^{4 \alpha + 3} ).
\end{equation}
Combing the results in (\ref{bd.I2})--(\ref{I20.bd}) and (\ref{I21.bd})--(\ref{I22.bd}), we have
\begin{equation}
\label{varate}
I_{n2}  =  O_P\Big(n^{-2} M^{4\alpha +3 } + n^{-1} M^{2\alpha +1}\Big).
\end{equation}
Combining the results in (\ref{triangle}),(\ref{biasrate}) and (\ref{varate}) and choosing $M \asymp n^{1/(2\alpha + 2\tau)}$, we obtain that 
\begin{eqnarray*}
	\|\widehat \beta - \beta_0\|^2 
	=O_P\big( n^{-2} M^{4\alpha +3 } + n^{-1}M^{2\alpha +1} + M^{-2\tau +1}\big) 
	=O_P\big(n^{- \frac{2 \tau - 1}{2\alpha + 2 \tau}}\big).
\end{eqnarray*}

\subsection{Proofs of Theorems~\ref{thm_covariance_discrete} and \ref{thm_beta_discrete}}
\label{ap.pf.partial}
{\bf Proof of Theorem~\ref{thm_covariance_discrete}.}  We begin with the $L_2$ rates of $\widetilde C_k$ for $k \geq 1.$ We wish to prove them in the same fashion as the proof of Theorem 1 in \cite{hansen2008}. For $p, q = 0,1,2,$ define
\begin{eqnarray*}
\widetilde{Z}_{p,q,i}^{(1)}(u,v) &=& \sum_{i=1}^{m_t} \sum_{j=1}^{m_{t+k}}K_{k,i,j,h,t}(u,v)\left ( \frac{U_{ti} - u}{h_C} \right)^p \left ( \frac{U_{(t+k)j} - v}{h_C} \right)^q,\\
\widetilde{Z}_{p,q,i}^{(2)}(u,v) &=& \sum_{i=1}^{m_t} \sum_{j=1}^{m_{t+k}}K_{k,i,j,h,t}(u,v)\left ( \frac{U_{ti} - u}{h_C} \right)^p \left ( \frac{U_{(t+k)j} - v}{h_C} \right)^q Z_{ti} Z_{(t+k)j}.
\end{eqnarray*}
Let $S_{pq}  =  (n \rho_n^2 h_C^2)^{-1}\sum_{i=1}^n \widetilde{Z}_{p,q,i}^{(1)}$ and $ G_{pq} = (n \rho_n^2 h_C^2)^{-1}\sum_{i=1}^n \widetilde{Z}_{p,q,i}^{(2)}.$ Then we have
$$
\widetilde C_k = \frac{(S_{20}S_{02} - S_{11}^2)G_{00} - (S_{10}S_{02} - S_{01}S_{11})G_{10} +(S_{10} S_{11}- S_{01}S_{20})G_{01}}{
(S_{20}S_{02} - S_{11}^2)S_{00} - (S_{10}S_{02} - S_{01}S_{11})S_{10} +(S_{10} S_{11}- S_{01}S_{20})S_{01}}
$$
so that $\widetilde{C}_k(u,v) - C_k(u,v)$ can be expressed as
\begin{eqnarray*}
\label{discrete_expression}
& = &\frac{(S_{20}S_{02} - S_{11}^2)\{G_{00} - C_k(u,v)S_{00} - h_C \frac{\partial C_k}{\partial u}(u,v)S_{10}- h_C\frac{\partial C_k}{\partial v}(u,v)S_{01} \}}{
(S_{20}S_{02} - S_{11}^2)S_{00} - (S_{10}S_{02} - S_{01}S_{11})S_{10} +(S_{10} S_{11}- S_{01}S_{20})S_{01}}\\
&-&
\frac{(S_{20}S_{02} - S_{11}^2)\{G_{10} - C_k(u,v)S_{10} - h_C \frac{\partial C_k}{\partial u}(u,v)S_{20}- h_C\frac{\partial C_k}{\partial v}(u,v)S_{11} \}}{
(S_{20}S_{02} - S_{11}^2)S_{00} - (S_{10}S_{02} - S_{01}S_{11})S_{10} +(S_{10} S_{11}- S_{01}S_{20})S_{01}}\\
&+&
\frac{(S_{20}S_{02} - S_{11}^2)\{G_{01} - C_k(u,v)S_{01} - h_C \frac{\partial C_k}{\partial u}(u,v)S_{11}- h_C\frac{\partial C_k}{\partial v}(u,v)S_{02} \}}{
(S_{20}S_{02} - S_{11}^2)S_{00} - (S_{10}S_{02} - S_{01}S_{11})S_{10} +(S_{10} S_{11}- S_{01}S_{20})S_{01}}.
\end{eqnarray*}
Let $\mathbb{U} = \{U_{ti},i=1,\ldots,m_t,t=1,\ldots,n\}.$ Suppose we have shown that for $p,q = 0,1,2,$
\begin{equation}
\label{numerator_rate}
\left\|G_{pq} - E \big\{G_{pq}\big|\mathbb{U}\big\}\right\|_{\cS} = O_P\left( \frac{1}{\sqrt{n \rho_n^2 h_C^2}} +\frac{1}{\sqrt{n}}\right ),
\end{equation}
and
\begin{equation}
\label{denominator_rate}
\sup_{u,v \in [0,1]}\big|S_{pq}(u,v) - ES_{pq}(u,v)\big| = o_P(1).
\end{equation}
By Taylor expansion, Condition~\ref{cond_2der} and (\ref{denominator_rate}),
\begin{equation}
\label{rate1}
 \left\|E \big\{G_{00}\big|\mathbb{U}\big\} - C_k(u,v)S_{00} - h_C \frac{\partial C_k}{\partial u}(u,v)S_{10}- h_C\frac{\partial C_k}{\partial v}(u,v)S_{01}\right\|_{\cS} = O_P(h^2_C).   
\end{equation}
Then combing (\ref{numerator_rate}) and (\ref{rate1}) yields that
\begin{equation}
\label{numerator_rate2}
\left\|G_{00} - C_k(u,v)S_{00} - h_C \frac{\partial C_k}{\partial u}(u,v)S_{10}- h_C\frac{\partial C_k}{\partial v}(u,v)S_{01}\right\|_{\cS} = O_P\left( \frac{1}{\sqrt{n \rho_n^2 h_C^2}} +\frac{1}{\sqrt{n}} + h_C^2\right ).
\end{equation}
Similarly, both $G_{10} - C_k(u,v)S_{10} - h_C \frac{\partial C_k}{\partial u}(u,v)S_{20}- h_C\frac{\partial C_k}{\partial v}(u,v)S_{11} $ and
$ G_{01} - C_k(u,v)S_{01} - h_C \frac{\partial C_k}{\partial u}(u,v)S_{11}- h_C\frac{\partial C_k}{\partial v}(u,v)S_{02} $  can be proved to have the same rate in (\ref{numerator_rate2}).
We can see from (\ref{denominator_rate}) that each denominator in $\widetilde{C}_k(u,v)$ is positive and bounded away from zero with probability approaching one, and as a consequence, part~(i) of Theorem~\ref{thm_covariance_discrete} follows.

Next, we turn to prove (\ref{numerator_rate}) and (\ref{denominator_rate}). For (\ref{numerator_rate}), it suffices to show that
 $$
 \int \int E\left\{G_{00}(u,v) -E \big\{G_{00}(u,v)\big|\mathbb{U}\big\}\right\}^2du dv \lesssim \frac{1}{n \rho_n^2 h^2_C} +\frac{1}{n},
 $$
 where  $a_n\lesssim b_n$ means $\limsup_{n\to \infty}|a_n/b_n| \le C$ for some positive constant $C>0.$ 
It is easy to see that 
\begin{eqnarray*}
E\left\{G_{00} - E \big\{G_{00}\big|\mathbb{U}\big\}\right\}^2
&\le & \frac{1}{n^2 \rho^4_n h_C^4}\sum_{t=1}^n E\left|\widetilde{Z}_{0,0,1}^{(2)} - E\big\{\widetilde{Z}_{0,0,1}^{(2)}|\mathbb{U}\big\}\right|^2 \\
&+& \frac{1}{n \rho^4_n h_C^4}\sum_{t= 1}^n \left|\cov\Big\{\widetilde{Z}_{0,0,1}^{(2)}-E\big\{\widetilde{Z}_{0,0,1}^{(2)}|\mathbb{U}\big\}, \widetilde{Z}_{0,0,t+1}^{(2)} - E\big\{\widetilde{Z}_{0,0,t+1}^{(2)}|\mathbb{U}\big\}\Big\}\right|.
\end{eqnarray*}
Let $\mathbb{W} = \{W_t(\cdot),U_{ti},i=1,\ldots,m_t,t=1,\ldots,n\}.$ Note that $\widetilde{Z}_{0,0,1}^{(2)} - E\big\{\widetilde{Z}_{0,0,1}^{(2)}|\mathbb{U}\big\} = \widetilde{Z}_{0,0,1}^{(2)} - E\big\{\widetilde{Z}_{0,0,1}^{(2)}\big | \mathbb{W}\big\} + E\big\{\widetilde{Z}_{0,0,1}^{(2)}\big | \mathbb{W} \big\} - E\big\{\widetilde{Z}_{0,0,1}^{(2)}|\mathbb{U}\big\}.$ Given $\mathbb{W},$ the first term $\widetilde{Z}_{0,0,1}^{(1)} - E\big\{\widetilde{Z}_{0,0,1}^{(1)}\big | \mathbb{W}\big\}$ is a U-type statistics and hence some specific calculations yield that
$
E\Big\{\widetilde{Z}_{0,0,1}^{(2)} - E\big\{\widetilde{Z}_{0,0,1}^{(2)}\big | \mathbb{W}\big\}\Big\}^2 \lesssim \rho_n^2h_C^2 + \rho_n^3h_C^3.
$
Moreover, $E\Big\{E\big\{\widetilde{Z}_{0,0,1}^{(2)}\big | \mathbb{W} \big\} - E\big\{\widetilde{Z}_{0,0,1}^{(2)}|\mathbb{U}\big\}\Big\}^2 \lesssim \rho_n^4h_C^4 + \rho_n^2h_C^2.$ As a result, 
$$
E|\widetilde{Z}_{0,0,1}^{(2)} - E\big\{\widetilde{Z}_{0,0,1}^{(2)}|\mathbb{U}\big\}|^2 \lesssim \rho_n^4h_C^4 +  \rho_n^2h_C^2.
$$
In a similar manner together with Marcinkiewicz-Zygmund inequality, we can show that
$$
E|\widetilde{Z}_{0,0,1}^{(2)}(u,v) - E\big\{\widetilde{Z}_{0,0,1}^{(2)}|\mathbb{U}\big\}|^s \lesssim \rho_n^{2s}h_C^{2s} + \rho_n^{s} h_C^{s} + \rho_n^{s/2}h_C^{2}.
$$

For each fixed $(u,v)$ and $h_C$, under Conditions~\ref{cond_discrete_mixing}--\ref{cond_observation_times}, we see that $(\widetilde Z_{0,0,1}^{(2)},\widetilde Z_{0,0,i}^{(2)},\ldots)$ is strictly stationary with $\psi$-mixing coefficients $\psi_Z(l)$ satisfying $\psi_Z(l) \lesssim (l - k)^{-\lambda}$ for $l \ge k+1.$ For $j^* < j \le \max(j^*+1,\rho_n^{-2} h_C^{-2})$ with fixed $j^* > k+1$, we have that
$$
\left|\cov\big\{\widetilde{Z}_{0,0,1}^{(2)}-E\big\{\widetilde{Z}_{0,0,1}^{(2)}|\mathbb{U}\big\}, \widetilde{Z}_{0,0,t+1}^{(2)} - E\big\{\widetilde{Z}_{0,0,t+1}^{(2)}|\mathbb{U}\big\}\big\}\right| \lesssim \rho_n^4h_C^4.
$$
For $j > \max(j^*+1,\rho_n^{-2} h_C^{-2}) + 1,$ using Davydov's lemma, we show that
$$
\left|\cov\big\{\widetilde{Z}_{0,0,1}^{(2)}-E\big\{\widetilde{Z}_{0,0,1}^{(2)}|\mathbb{U}\big\}, \widetilde{Z}_{0,0,t+1}^{(2)}-E\big\{\widetilde{Z}_{0,0,t+1}^{(2)}|\mathbb{U}\big\}\big\}\right| \lesssim j^{-2 + 2/s} \big( \rho_n^4 h_C^4 + \rho_n^2 h_C^2 + \rho_n h_C^{4/s}\big).
$$
Therefore, the rate in (\ref{numerator_rate}) follows from the steps to prove Theorem~1 in \cite{hansen2008}. Similarly, together with Conditions~\ref{cond_discrete_mixing}--\ref{cond_bandwidths}, the rates in (\ref{denominator_rate}) and $\|\widehat{S}_k - S_k\|$ follows from the steps to prove Theorem~2 in \cite{hansen2008}. The proof is complete.\\
{\bf Proof of Theorem~\ref{thm_beta_discrete}.} By Theorem~\ref{thm_covariance_discrete} for $k=1,\dots, L,$ we can easily show that 
\begin{equation}
\label{rate_K.R}
    \|\widetilde K - K\|_{\cS} = O_P\big( \delta_{n1}\big) ~\mbox{and}~~
\|\widetilde R - R\| = O_P\big( \delta_{n1} + \delta_{n2}\big).
\end{equation}
Following directly from the proof steps of Theorem~\ref{thm_est_scalar} by replacing $\|\widehat{K} - K\|_{\cS}=O_P(n^{-1/2})$ and $\|\widehat{R} - R\|=O_P(n^{-1/2})$ with the corresponding rates in (\ref{rate_K.R}), we complete our proof.

\linespread{1.03}\selectfont
\bibliography{paperbib}
\bibliographystyle{dcu}

\newpage
\linespread{1.59}\selectfont
\begin{center}
	{\noindent \bf \large Supplementary Material to ``Functional Linear Regression: Dependence and Error Contamination"}\\
\end{center}
\begin{center}
	{\noindent Cheng Chen, Shaojun Guo and Xinghao Qiao}
\end{center}

\bigskip
This supplementary material contains proofs of Theorem~\ref{thm_est_func} and all technical lemmas in Appendix~\ref{sec.addproof}, the presentation of the basis expansion approach to address partially observed curve time series in Appendix~\ref{sec.basis.app} and additional simulation results in Appendix~\ref{supp.sec.emp}.
\setcounter{page}{1}
\setcounter{section}{1}
\setcounter{equation}{0}
\renewcommand{\theequation}{A.\arabic{equation}}

\section{Additional technical proofs}
\label{sec.addproof}
\subsection{Proof of Theorem~\ref{thm_est_func}}
\label{pf.thm2}

Following the similar arguments used in the proofs for Lemmas~\ref{lemma.est.R} and \ref{lemma.R.bd} under some regularity conditions, we can show that 
\begin{equation}
\label{H_rate}
\|\widehat H - H \|_{\cS} = O_P(n^{-1/2})~\mbox{and}~ \|H\|_{\cS} = O(1).
\end{equation}
Consider the case when $d$ is fixed. Let $\widetilde\gamma(u,v)=\int_{\cU}\widecheck K^{-1}(u,w) \widehat H(w,v)dw.$ Then we have
\begin{equation}
\label{gamma.bd}
\|\widetilde\gamma-\gamma_0\|_{\cS} \leq \|\widecheck K^{-1}-K^{-1}\|_{\cS}\|\widehat H\|_{\cS} + \|K^{-1}\|_{\cS}\|\widehat H-H\|_{\cS}.    
\end{equation}
It follows from Lemma~\ref{lemma.Kinv} and (\ref{H_rate}) that  $\|\widetilde\gamma-\gamma\|_{\cS} = O_P(n^{-1/2}+n^{-1/2})=O_P(n^{-1/2}).$ Finally, applying the similar technique used in the proof for part~(i) of Theorem~\ref{thm_est_scalar}, we can prove the result in part~(i) of Theorem~\ref{thm_est_func}.

When $d=\infty,$ let $\gamma_M(u,v)=\sum_{j=1}^{M} \theta_j^{-1} \psi_j(u) \langle \psi_j, H(\cdot,v) \rangle.$ By the triangle inequality, we have 
\begin{eqnarray}
\label{triangle2}
\|\widehat \gamma -\gamma_0\|^2_{\cS} \le \|\widehat \gamma-\gamma_M\|^2_{\cS} + \|\gamma_M-\gamma_0\|^2_{\cS}.
\end{eqnarray}
It follows from Condition~\ref{cond_bias_func} and some specific calculations that 
\begin{eqnarray}
\label{biasrate.func}
\|\gamma_M - \gamma_0\|_{\cS}^2
& =& O(1)\Big\|\sum_{j=M+1}^{\infty}\sum_{\ell =1}^\infty b_{j \ell}\psi_j(u)\psi_{\ell}(v)\Big\|_{\cS}^2\nonumber\\
&=& O(1)\sum_{j=M+1}^{\infty}\sum_{\ell =1}^\infty b_{j \ell}^2 = O(1) \sum_{j=M+1}^{\infty}\sum_{\ell =1}^\infty(j+\ell)^{-2\tau-1} = O(M^{-2\tau + 1}).
\end{eqnarray}
It remains to show that the convergence rate of $\|\widehat \gamma-\gamma_M\|^2_{\cS}$.
Observe that 
\begin{eqnarray*}
	\widehat \gamma(u,v) - \gamma_M(u,v) 
	&=& \sum_{j=1}^{M} \big (\widehat \theta_j^{-1} - \theta_j^{-1} \big)\langle \psi_j,H \rangle(v) \widehat \psi_j(u) \\
	&&  + \sum_{j=1}^{M} \widehat \theta_j^{-1} \big(\langle \widehat \psi_j, \widehat H \rangle(v) - \langle \psi_j, H \rangle\big)(v) \widehat \psi_j(u) \\
	&&  + \sum_{j=1}^{M} \theta_j^{-1} \langle  \psi_j, H \rangle(v) \big\{\widehat\psi_j(u) - \psi_j(u)\big\}.
\end{eqnarray*}
Then we have,
\begin{eqnarray*}
	\|\widehat \gamma - \gamma_M\|^2_\cS
	&\le& 3\sum_{j=1}^{M} \big (\widehat \theta_j^{-1} - \theta_j^{-1} \big)^2 \|\langle \psi_j,H \rangle\|^2 + 3\sum_{j=1}^{M} \widehat \theta_j^{-2} \big\|\langle \widehat \psi_j, \widehat H \rangle - \langle \psi_j, H \rangle\big\|^2 \nonumber \\
	&& \hskip 0.5cm + ~3 M\sum_{j=1}^{M} \theta_j^{-2} \big\|\langle  \psi_j, H \rangle\big\|^2 \big\|\widehat\psi_j - \psi_j\big\|^2. \nonumber 
\end{eqnarray*}
Following the similar arguments used in the proof for Theorem~\ref{thm_est_scalar} (ii),  we can show that 
\begin{eqnarray}
\label{varrate.func}
\|\widehat \gamma-\gamma_M\|^2_{\cS} = O_P(M^{4\alpha +3} n^{-2} + M^{2\alpha +1} n^{-1}).
\end{eqnarray} 
Combing the results in (\ref{triangle2})--(\ref{varrate.func}) and choosing $M \asymp n^{1/ (2\alpha +2 \tau)}$, we have 
$$
\|\widehat\gamma-\gamma_0\|_{\cS}^2 = O_P\big( M^{2\alpha +1}n^{-1} + M^{-2\tau +1}\big) 
=O_P\big(n^{- \frac{2 \tau - 1}{2\alpha + 2 \tau}}\big).
$$
which completes our proof for part~(ii) of Theorem~\ref{thm_est_func}.

\subsection{Lemma~\ref{lemma.asy.eigen} and its proof}
\begin{lemma}
	\label{lemma.asy.eigen}
	Suppose that Conditions~\ref{cond_mix}--\ref{cond_eigen} hold and $\langle \widehat \psi_j, \psi_j \rangle \ge 0$. Then as $n \rightarrow \infty,$ the following results hold:\\
	(i) $\|\widehat K-K\|_{\cS}=O_P(n^{-1/2})$ and  $\sup_{j \ge 1}|\widehat\theta_j - \theta_j|=O_P(n^{-1/2}).$ \\  
	(ii) When $d$ is fixed, $\|\widehat\psi_j-\psi_j\|=O_P(n^{-1/2})$ for $j =1,\ldots,d.$\\
	(iii)When $d = \infty$, $\|\widehat\psi_j-\psi_j\|=O_P(j^{1+\alpha}n^{-1/2})$ for $j =1,2,\ldots.$
\end{lemma}
{\bf Proof.} The first result in part~(i) can be found in Theorem~1 of \textcolor{blue}{Bathia et al. (2010)} and hence the proof is omitted. By (4.43) of \textcolor{blue}{Bosq (2000)}, we have ${\sup}_{j \geq 1} |\widehat\theta_j - \theta_j| \leq \|\widehat K-K\|_{\cS}=O_P(n^{-1/2}),$ which completes the proof for the second result in part~(i). To prove parts~(ii) and (iii), let $\delta_{j}=2\sqrt{2}\max\{(\theta_{j-1}-\theta_j)^{-1},(\theta_{j}-\theta_{j+1})^{-1}\}$ if $j \geq 2$ and $\delta_1=2\sqrt{2}(\theta_1-\theta_2)^{-1}.$ It follows from Lemma~4.3 of \textcolor{blue}{Bosq (2000)} that $\|\widehat\psi_j-\psi_j\| \leq \delta_j \|\widehat K-K\|_{\cS} =O_P(\delta_j n^{-1/2}).$ Under Condition~\ref{cond_eigen}(i) with a fixed $d,$ root-$n$ rate can be achieved. When  $d=\infty,$ Condition~\ref{cond_eigen}(ii) and (iii) imply that $\delta_j \leq Cj^{\alpha+1}$  with some positive constant $C.$ 
This completes our proof for part~(iii).

\subsection{Lemma~\ref{lemma.est.R} and its proof}
\begin{lemma}
	\label{lemma.est.R}
	Suppose that Conditions~\ref{cond_mix}-\ref{cond_moment_scalar} hold, then 
	$\|\widehat R- R\|=O_P(n^{-1/2}).$
\end{lemma}
{\bf Proof.} Provided $L$ is fixed, we may set $n \equiv n-L.$ Let $\cS$ denotes the space consisting of all the operators with a finite Hilbert-Schmidt norm and $\cH$ denotes the space consisting of all the functions with a finite $L_2$ norm. Let $Z_{tk}=W_t \otimes W_{t+k} \in {\cS}$ and $z_{tk}=Y_t W_{t+k} \in {\cH}.$ Now consider the kernel $\rho:$ $\cS \times \cH \rightarrow \cH$ given by $\rho(A,x)=Ax^*$ with $A \in \cS$ and $x \in \cH.$ Let $c_k(\cdot)=\cov\{Y_t, W_{t+k}(\cdot)\}.$ We can represent $\widehat C_k \widehat c_k^*= n^{-2} \sum_{t=1}^n\sum_{t'=1}^n \rho(Z_{tk},z_{t'k}),$ which is simply a $\cH$ valued Von Mises' functional (\textcolor{blue}{Borovskikh, 1996}). For $d \geq 1,$ neither of $C_k$ and $c_k$ is zero, it follows from Lemma~3 of \textcolor{blue}{Bathia et al. (2010)} that $E\|\widehat C_k \widehat c_k^*-C_k c_k^*\|^2=O(n^{-1}).$ Then by the Chebyshev inequality, we have $$\|\widehat R-R\| \leq \sum_{k=1}^L\|\widehat C_k \widehat c_k^*-C_k c_k^*\|=O_P(n^{-1/2}),$$ which completes the proof.

\subsection{Lemma~\ref{lemma.R.bd} and its proof}
\begin{lemma} 
	\label{lemma.R.bd}
	Suppose that Condition~\ref{cond_moment_scalar} holds, then $\|R\|=O(1).$
\end{lemma}
{\bf Proof.} By the definitions of $C_k$ and (\ref{def.R}), we have 
$$\|R\| \leq \sum_{k=1}^L \|C_k\|_{\cS}\|\cov(Y_t,W_{t+k})\|=\sum_{k=1}^L\|E\{W_t(u)W_{t+k}(v)\}\|_{\cS}\|E(Y_tW_{t+k}(u))\|.$$ It follows from Cauchy-Schwartz inequality, Condition~\ref{cond_moment_scalar}, Fubini Theorem and Jensen's inequality that $\|E\{W_t(u)W_{t+k}(v)\}\|_{\cS}^2$
\begin{eqnarray*}
	&=&\int_{\cU}\int_{\cU}[E\{W_t(u)W_{t+k}(v)\}]^2dudv \\
	&\leq& \int_{\cU}E\{W_t(u)^2\}du\int_{\cU}E\{W_{t+k}(v)^2\}dv=\Big[\int_{\cU}E \{W_t(u)^2\}du\Big]^2 \leq E\Big\{\int_{\cU} W_t(u)^2du\Big\}^2 <\infty.
\end{eqnarray*}
Similarly, $\|E\{Y_tW_{t+k}(u)\}\|^2 \leq E(Y_t^2)\int_{\cU}E\{W_{t+k}(u)^2\}du<\infty.$ Combining the above results leads to $\|R\|=O(1).$

\subsection{Lemma~\ref{lemma.d} and its proof}
\begin{lemma}
	\label{lemma.d}
	Suppose the Conditions~\ref{cond_mix}, \ref{cond_moment_scalar},  \ref{cond_eigen} (i) and (iii) hold. Let $\epsilon_n \rightarrow 0,$ $\epsilon_n^2 n \rightarrow \infty$ and as $n\rightarrow \infty.$ Then when $d < \infty$, $P\big(\widehat d \neq d\big)=O\{(\epsilon_n^2 n)^{-1}\}\rightarrow 0.$
\end{lemma}
{\bf Proof}. This lemma, which holds for $d<\infty,$ can be found in Theorem~3 of \textcolor{blue}{Bathia et al. (2010)} and hence the proof is omitted.

\subsection{Lemma~\ref{lemma.Kinv} and its proof}
\begin{lemma}
	\label{lemma.Kinv} 
	Suppose that Conditions~\ref{cond_mix}, \ref{cond_moment_scalar}, \ref{cond_eigen}(i) and (iii) hold. Then the following results hold.\\
	(i) $\|\widecheck K^{-1}-K^{-1}\|_{\cS}=O_P(n^{-1/2}).$\\
	(ii) $\|K^{-1}\|_{\cS}=O(1).$
\end{lemma}
{\bf Proof}. Observe that
$$
\widecheck K^{-1}-K^{-1} = 
\sum_{j=1}^d (\widehat\theta_j^{-1}-\theta_j^{-1})\widehat\psi_j(u)\widehat\psi_j(v) + \sum_{j=1}^d \theta_j^{-1}
\{\widehat\psi_j(u)\widehat\psi_j(v)-\psi_j(u)\psi_j(v)\}.
$$ 
Then by the orthonormality of $\{\psi_j(\cdot)\}$ and $\{\widehat\psi_j(\cdot)\},$  we have
\begin{eqnarray}
\|\widecheck K^{-1} -K^{-1}\|_{\cS} 
\leq \sum_{j=1}^d \widehat \theta_j^{-1}\theta_j^{-1}|\widehat \theta_j-\theta_j| + 2\sum_{j=1}^d \theta_j^{-1} \|\widehat \psi_j-\psi_j\|. \label{K.bd}
\end{eqnarray}
When $d$ is fixed, the smallest eigenvalue $\theta_d$ is bounded away from zero. It follows from Lemma~\ref{lemma.asy.eigen} (i),(ii) and (\ref{K.bd}) that there exists some positive constant C such that $\|\widecheck K^{-1} -K^{-1}\|_{\cS} \leq C (\theta_d^{-2}+\theta_d^{-1}) n^{-1/2},$ which completes the proof for part~(i).

Note that $\|K^{-1}\|_{\cS}=\|\sum_{j=1}^d \theta_j^{-1}\psi_j(u)\psi_j(v)\|_{\cS}=(\sum_{j=1}^d \theta_j^{-2})^{1/2}\leq d^{1/2}\theta_d^{-1}$. Then part~(ii) follows as $d$ is fixed and $\theta_d$ is bounded below from zero.

\subsection{Lemma~\ref{eigen.expansion} and its proof}
\begin{lemma}
	\label{eigen.expansion}
	If $\inf_{k \neq j}|\widehat \theta_j - \theta_k| > 0$, then
	\begin{eqnarray}
	\widehat \psi_j - \psi_j = \sum_{k: k \neq j } (\widehat \theta_j - \theta_k)^{-1} \psi_k \langle \hat \psi_j,  \langle \widehat K - K, \psi_k\rangle \rangle + \psi_j  \langle \widehat \psi_j - \psi_j, \psi_j \rangle. 
	\end{eqnarray}
\end{lemma}
{\bf Proof.} This lemma can be derived from Lemma 5.1 of \textcolor{blue}{Hall and Horowitz (2007)} and hence the proof is omitted.

\section{Basis expansion approach}
\label{sec.basis.app}
We develop a standard basis expansion approach to estimate $K(u,v)$ and $R(u).$ 
Let $\bB(u)$ be the $J$-dimensional orthonormal basis function, i.e. $\int_{\cU} \bB(u) \bB^{\T}(u)du = \bI_{J},$ such that each $C_k(u,v)$ can be well approximated by $\{\bB(u)\}^T\bSigma_k\bB(v)$. In practice, $J$ can be selected by a similar cross-validation procedure described in Section~\ref{sec.tune}. Let $\bB_{ti} = \bB(U_{ti}).$ We consider minimizing 
\begin{equation}
\label{incomp_crit}
  \sum_{t = 1 }^{n - L} \sum_{i=1}^{m_t}\sum_{j=1}^{m_{t+k}} \left\{ Z_{ti} Z_{(t+k)j} - \bB_{ti}^T \bSigma_k \bB_{(t+k)j}\right\}^2  
\end{equation}
with respect to $\bSigma_{k} \in \mR^{J \times J}.$ Standard calculation shows that the estimate of $\bSigma_k$ that minimizes (\ref{incomp_crit}) is 
    \label{incomp_sol}
$$   \vect(\widehat\bSigma_k) = \left(\sum_{t,i,j} (\bB_{(t+k)j} \otimes \bB_i)(\bB_{(t+k)j} \otimes \bB_i)^T\right)^{-1} \sum_{t,i,j}(\bB_{(t+k)j} \otimes \bB_i)Z_{ti} Z_{(t+k)j},$$
where $\vect(\bB)$ denotes the vectorization of the matrix $\bB$ formed by stacking its columns into a single column vector and $\otimes$ is the Kronecker product. Then the estimate of $K(u,v)$ is $$\widetilde{K}(u,v) = \{\bB(u)\}^{\T}\sum_{k=1}^L\widehat{\bSigma}_k\widehat{\bSigma}_k^{\T}\bB(v).$$


Similarly, we can obtain a consistent estimator $\widehat{\cov}\{Y_t,W_{t+k}(u)\} = \widehat{\bdelta}_k^T\bB(u)$, where $\widehat{\bdelta}_k$ is obtained by minimizing 
$$
\sum_{t = 1 }^{n - L} \sum_{1 \le i \le m_t} \left\{ Y_t Z_{(t+k)i}  - \bdelta_k^T\bB_{(t+k)i} \right\}^2
$$
with repsect to $\bdelta_k \in {\mathbb R}^J.$ Then the estimate of $\bdelta_k$ is
$$
\widehat{\bdelta}_k = \left(\sum_{t,i} \bB_{(t+k)i}\bB_{(t+k)i}^T \right)^{-1} \sum_{t,i} \bB_{(t+k)i}Y_t Z_{(t+k)i}.
$$
As a result, $R(u)$ can be estimated by 
$$
\widetilde{R}(u) = \sum_{k=1}^L\widehat{\bdelta}_k^T\widehat{\bSigma}_k^T\bB(u).
$$

\section{Additional simulation results}
\label{supp.sec.emp}

For Example~2, Table~\ref{table.m2.eigen} reports the variance explained by each of the $10$ components under the population level. For each of the three parts corresponding to $d=2, 4$ and $6$, the second and third rows provide the variance explained by each of the $d$ signal components and $10$ error components, respectively. The first row ranks the components based on the overall variance explained by each individual component, where the fourth row displays the corresponding values. Take $d=4$ as an illustrative example, the autocovariance-based approach can correctly identify the first four signal components, while the covariance-based approach can only correctly identify ``1" and ``2", but incorrectly select ``7" and ``8" as signal components. Moreover, we consider another scenario for Example~2 by generating innovations $\{\nu_{tj}\}$ from a standard normal distribution, where the variance decomposition is illustrated via Table~\ref{table.m2.eigen.old}. Under this setting, we can observe that both approaches are capable of correctly identifying the $d$ signal components. 

We next illustrate the sample performance of AGMM using two additional simulated examples to support Section~\ref{subsec.sim}.

\begin{table}
	\caption{\label{table.m2.eigen} The variance explained by each of the components in Example~2. Top $d$ components identified by covaraicne-based and autocovariance-based approaches are underlined and in bold font, respectively.}
	\begin{center}
		\resizebox{5.7in}{!}{
			\begin{tabular}{cccccccccccc}
				
				\hline
				
				\multirow{5}{*}{d=2}
				
				& Component  & 1    & 2    & 7    & 8     & 9     & 10    & 3     & 4     & 5     & 6     \\
				
				&Signal & {\bf1.73} & {\bf1.19} &      &       &       &       &       &       &       &       \\
				
				&Error & 1.00 & 0.50 & 1.57 & 1.49  & 1.40  & 1.32  & 0.25  & 0.13  & 0.06  & 0.03  \\
				
				&Sum & \underline{2.73} & \underline{1.69} & 1.57 & 1.49  & 1.40  & 1.32  & 0.25  & 0.13  & 0.06  & 0.03  \\

				
				\hline
				
				\multirow{5}{*}{d=4}
				
				& Component & 1    & 2    & 7    & 8     & 9     & 10    & 3     & 4     & 5     & 6     \\
				
				&Signal & {\bf2.50} & {\bf1.73} &      &       &       &       & {\bf1.38}  & {\bf1.19}  &       &       \\
				
				&Error & 1.00 & 0.50 & 1.73 & 1.64  & 1.55  & 1.45  & 0.25  & 0.13  & 0.06  & 0.03  \\
				
				&Sum & \underline{3.50} & \underline{2.23} & \underline{1.73} & \underline{1.64}  & 1.55  & 1.45  & 1.63  & 1.32  & 0.06  & 0.03  \\

				
				\hline
				
				\multirow{5}{*}{d=6}
				
				& Component  & 1    & 2    & 3    & 7     & 8     & 9     & 10    & 4     & 5     & 6     \\
				
				&Signal & {\bf3.00} &{\bf 2.16} & {\bf1.73} &       &       &       &       & {\bf1.47}  & {\bf1.30}  & {\bf1.19}  \\
				
				&Error & 1.00 & 0.50 & 0.25 & 1.90  & 1.80  & 1.70  & 1.60  & 0.13  & 0.06  & 0.03  \\
				
				&Sum & \underline{4.00} & \underline{2.66} & \underline{1.98} & \underline{1.90}  & \underline{1.80}  & \underline{1.70}  & 1.60  & 1.60  & 1.37  & 1.22  \\
				
				
				\hline
				
			\end{tabular}
		}	
	\end{center}
\end{table}

\begin{table}
	\caption{\label{table.m2.eigen.old} The variance explained by each of the components in Example~2 with $\{\nu_{tj}\}$ being $N(0,1)$ variables. Top $d$ components identified by covaraicne-based and autocovariance-based approaches are underlined and in bold font, respectively.}
	\begin{center}
		\resizebox{5.7in}{!}{
			\begin{tabular}{cccccccccccc}
				\hline
				\multirow{6}{*}{d=2}
				& Component    & 1   & 2    & 3    & 4     & 5     & 6     & 7     & 8     & 9     & 10    \\
				& Signal   & {\bf1.73} & {\bf1.19} &      &       &       &       &       &       &       &       \\
				& Error   & 1.00 & 1.00 & 1.00 & 1.00  & 1.00  & 1.00  & 1.00  & 1.00  & 1.00  & 1.00  \\
				& Sum  & \underline{2.73} & \underline{2.19} & 1.00 & 1.00  & 1.00  & 1.00  & 1.00  & 1.00  & 1.00  & 1.00  \\
				\hline
				\multirow{6}{*}{d=4}
				& Component    & 1 & 2 & 3 & 4  & 5  & 6  & 7  & 8  & 9  & 10 \\
				& Signal   & {\bf2.50} & {\bf1.73} & {\bf1.38} &{\bf 1.19}  &       &       &       &       &       &       \\
				& Error  & 1.00 & 1.00 & 1.00 & 1.00  & 1.00  & 1.00  & 1.00  & 1.00  & 1.00  & 1.00  \\
				& Sum  & \underline{3.50} & \underline{2.73} & \underline{2.38} & \underline{2.19}  & 1.00  & 1.00  & 1.00  & 1.00  & 1.00  & 1.00  \\
				\hline
				\multirow{6}{*}{d=6}
				& Component   & 1    & 2    & 3    & 4    & 5     & 6     & 7     & 8     & 9     & 10    \\
				& Signal   & {\bf3.00} & {\bf2.16} & {\bf1.73} & {\bf1.47}  & {\bf1.30}  & {\bf1.19}  &       &       &       &       \\
				& Error  & 1.00 & 1.00 & 1.00 & 1.00  & 1.00  & 1.00  & 1.00  & 1.00  & 1.00  & 1.00  \\
				& Sum  & \underline{4.00} & \underline{3.16} & \underline{2.73} & \underline{2.47}  & \underline{2.30}  & \underline{2.19}  & 1.00  & 1.00  & 1.00  & 1.00  \\
				\hline
			\end{tabular}
		}	
	\end{center}
\end{table}

\textbf{Example~4}:
This example is used to demonstrate the superiority of AGMM methods under the setting where the dimension of the $\beta_0(\cdot)$ is less than the dimension of $X_t(\cdot).$ While the data are generated in the same fashion to Example 2, the slope functions are generated by $\beta_0(\cdot)=\sum_{j=1}^d \widetilde b_j \phi_j(\cdot)$ with $\widetilde b_j=b_j$ for $j=1, \dots, d-1$ and $\widetilde b_d=0$ so that the dimension of $\beta_0(\cdot)$ is $d-1<d.$  Table~\ref{table.m4} provides numerical results under the oracle scenario with true $d$ in the estimation. We obtain the same findings to those in Table~\ref{table.m2}, i.e. two versions of AGMM significantly outperform their competing methods, while AGMM is superior to Base AGMM in most of the cases.


\begin{table}[ht]
	\caption{\label{table.m4} {\it Example~4}: The mean and standard error (in parentheses) of the mean integrated squared error for $\widehat \beta(u)$ over 100 simulation runs. The lowest values are in bold font.}
	\begin{center}
		\resizebox{6.2in}{!}{
			\begin{tabular}{cc|cccccc}
				$n$ & $d$ & Base CLS & CLS & Base CGMM & Base ALS & Base AGMM & AGMM \tabularnewline
				\hline
				\multirow{3}{*}{400}
				& 2 & 0.683(0.008) & 0.577(0.007) & 0.244(0.014) & 0.646(0.008) & 0.255(0.014) & {\bf 0.132(0.008)} \tabularnewline
				& 4 & 1.415(0.042) & 0.993(0.043) & 1.619(0.072) & 0.756(0.014) & 0.489(0.024) & {\bf 0.324(0.018)} \tabularnewline
				& 6 & 1.990(0.051) & 1.600(0.055) & 2.312(0.066) & 0.775(0.021) & 0.647(0.025) & {\bf 0.500(0.024)} \tabularnewline
				\multirow{3}{*}{800}
				& 2 & 0.589(0.006) & 0.560(0.006) & 0.137(0.008) & 0.593(0.006) & 0.125(0.005) & {\bf 0.076(0.005)} \tabularnewline
				& 4 & 1.378(0.038) & 0.855(0.037) & 1.641(0.069) & 0.620(0.008) & 0.253(0.010) & {\bf 0.191(0.012)} \tabularnewline
				& 6 & 1.817(0.036) & 1.546(0.035) & 2.351(0.077) & 0.515(0.009) & {\bf 0.295(0.011)} & 0.304(0.019) \tabularnewline
				\multirow{3}{*}{1200}
				& 2 & 0.573(0.004) & 0.552(0.004) & 0.081(0.005) & 0.576(0.004) & 0.082(0.004) & {\bf 0.048(0.003)} \tabularnewline
				& 4 & 1.383(0.035) & 0.875(0.044) & 1.732(0.072) & 0.554(0.005) & 0.142(0.006) & {\bf 0.108(0.006)} \tabularnewline
				& 6 & 1.895(0.032) & 1.623(0.036) & 2.598(0.071) & 0.462(0.007) & {\bf 0.196(0.007)} & 0.197(0.013) \tabularnewline
				\hline
				\vspace{-0.8cm}
			\end{tabular}
		}	
	\end{center}
\end{table}

{\bf Example~5}: 
This example is used to illustrate the advantages of AGMM methods under the infinite dimensional setting. With a large enough $d,$ e.g. $d=25,$ the data is generated as follows so that Conditions~\ref{cond_eigen} and \ref{cond_bias_scalar} are satisfied. To be specific, we generate $X_t(u) = \sum_{j=1}^{d} \xi_{tj}\phi_j(u)$ based on $\xi_{tj} = 0.8\xi_{t-1,j} + \epsilon_{tj},$ where $\epsilon_{tj}\sim N(0, j^{-0.75}).$ Some specific calculations yield lag-$k$ autocovariance of $\xi_{tj}$ as $\cov(\xi_{tj}, \xi_{t+k,j}) = \frac{0.8^k \cdot j^{-0.75}}{0.36}$ and eigenvalues of $K$ in equation (\ref{def.K}) as $\theta_j = \sum_{k=1}^L \cov(\xi_{tj}, \xi_{t+k,j})^2 = \frac{\sum_{k=1}^L 0.8^{2k}}{0.36^2}\cdot j^{-1.5} \asymp j^{-1.5}$ under the orthonormality of $\{\phi_j(\cdot)\}_{j \geq 1}.$ Hence, Condition \ref{cond_eigen} is satisfied with $\alpha = 1.5.$ Moreover, we set $\tau = 2$ in Condition \ref{cond_bias_scalar} so that $\tau \geq \alpha+1/2$ is satisfied and hence generate the slope function $\beta_0(\cdot)=\sum_{j=1}^{d} \widetilde b_j \phi_j(\cdot)$ with $\widetilde b_j = (-1)^{j-1}\cdot 2\cdot j^{-2}.$ The innovations $\{\nu_{tj}\}_{n \times 10}$ are independent $N(0,1)$ variables. The truncated dimension $M$ is chosen so that the top $M$ eigenvalues explains over 90\% of the total variation. Table~\ref{table.m5} reports numerical results for all comparison methods under two settings, where $\{\phi_j(\cdot)\}_{j=1}^d$ and $\{\zeta_j(\cdot)\}_{j=1}^{10}$ are generated from the corresponding basis functions used in Example~1 and 2, respectively. Again we observe the prominent superiority of two versions of AGMM methods over the competitors with AGMM significantly outperforming Base AGMM.

\begin{table}[ht]
	\caption{\label{table.m5} {\it Example~5}: The mean and standard error (in parentheses) of the mean integrated squared error for $\widehat \beta(u)$ over 100 simulation runs. The lowest values are in bold font.}
	\begin{center}
		\resizebox{6.5in}{!}{
			\begin{tabular}{cc|cccccc}
				$\{\phi_j\}_{j=1}^{25}, \{\zeta_j\}_{j=1}^{10}$ & $n$ & Base CLS & CLS & Base CGMM & Base ALS & Base AGMM & AGMM \tabularnewline
				\hline
				\multirow{3}{*}{Example 1}
				& 400 &  0.972(0.017) & 1.068(0.022) & 0.913(0.030) & 0.708(0.012) & 0.582(0.013) & {\bf 0.390(0.017)} \tabularnewline
			    & 800 &  0.810(0.011) & 0.849(0.012) & 0.540(0.018) & 0.535(0.008) & 0.329(0.008) & {\bf 0.200(0.008)} \tabularnewline
				& 1200 & 0.775(0.009) & 0.800(0.009) & 0.446(0.017) & 0.463(0.006) & 0.235(0.005) & {\bf 0.156(0.005)} \tabularnewline
				\multirow{3}{*}{Example 2}
				& 400 &  0.677(0.012) & 0.684(0.015) & 0.838(0.026) & 0.702(0.012) & 0.590(0.014) & {\bf 0.376(0.017)} \tabularnewline
			    & 800 &  0.536(0.008) & 0.541(0.007) & 0.449(0.012) & 0.546(0.007) & 0.341(0.008) & {\bf 0.200(0.007)} \tabularnewline
				& 1200 & 0.482(0.005) & 0.486(0.005) & 0.308(0.009) & 0.478(0.004) & 0.241(0.005) & {\bf 0.153(0.005)} \tabularnewline
				\hline
			\end{tabular}
			\vspace{-0.8cm}
		}	
	\end{center}
\end{table}

\linespread{1.1}\selectfont
\section*{References}
\begin{description}
	\item Bathia, N., Yao, Q.  and Ziegelmann, F. (2010). Identifying the finite dimensionality of curve time series. {\it The Annals of Statistics}, {\bf 38}, 3352-3386.
	\item 
	Borovskik, Y. V. (1996). {\it U Statistics in Banach Spaces}. VSP, Netherlands.
    \item 
	Bosq, D. (2000). {\it Linear Processes in Function Spaces - Theory and Applications}. Springer, New York.
	\item
	Hall, P.  and Horowitz, J. L. (2007). Methodology and convergence rates for functional linear regression. {\it The Annals of Statistics}, {\bf 35}, 70-91.
\end{description}
\end{document}